\documentclass{article}
\usepackage{arxiv}
\pdfoutput=1
\usepackage[utf8]{inputenc} 
\usepackage[T1]{fontenc}    
\usepackage{hyperref}       
\usepackage{url}            
\usepackage{booktabs}       
\usepackage{amsfonts}       
\usepackage{nicefrac}       
\usepackage{microtype}      
\usepackage{graphicx}
\usepackage{subcaption}
\usepackage{multirow}
\usepackage{amsmath}
\makeatletter
\newcommand\footnoteref[1]{\protected@xdef\@thefnmark{\ref{#1}}\@footnotemark}
\makeatother

\title{DARTS: DenseUnet-based Automatic Rapid Tool for brain Segmentation}
\author{
  Aakash Kaku $^\dagger$\thanks{Equal contribution} \\
  \texttt{ark5765@nyu.edu} \\
  \And
  Chaitra V. Hegde \thanks{Center for Data Science, New York University,New York, NY 10011} $^*$ \\
  \texttt{cvh255@nyu.edu} \\
  \And
  Jeffrey Huang \thanks{Department of Radiology, New York University School of Medicine,New York, NY 10016} \\
  \texttt{Jeffrey.Huang@nyulangone.org} \\
  \And
  Sohae Chung $^\ddagger$ \\
  \texttt{Sohae.Chung@nyulangone.org} \\
  \And
  Xiuyuan Wang $^\ddagger$\\
  \texttt{Xiuyuan.Wang@nyulangone.org} \\
  \And
  Matthew Young $^\ddagger$ \\
  \texttt{Matthew.Young3@nyulangone.org} \\
  \And
  Alireza Radmanesh $^\ddagger$ \\
  \texttt{Alireza.Radmanesh@nyulangone.org} \\
  \And
  Yvonne W. Lui $^\ddagger$ $^\S$   \\
  \texttt{Yvonne.Lui@nyulangone.org} \\
  \And
  Narges ~Razavian $^{\dagger \ddagger}$ \thanks{Co-corresponding authors}\\
  \texttt{Narges.Razavian@nyulangone.org} \\
}
\begin{document}

\maketitle

\begin{abstract}
Quantitative, volumetric analysis of Magnetic Resonance Imaging (MRI) is a fundamental way researchers study the brain in a host of neurological conditions including normal maturation and aging. Despite the availability of open-source brain segmentation software, widespread clinical adoption of volumetric analysis has been hindered due to processing times and reliance on manual corrections. 
Here, we extend the use of deep learning models from proof-of-concept, as previously reported, to present a comprehensive segmentation of cortical and deep gray matter brain structures matching the standard regions of aseg+aparc included in the commonly used open-source tool, Freesurfer. The work presented here provides a real-life, rapid deep learning-based brain segmentation tool to enable clinical translation as well as research application of quantitative brain segmentation. The advantages of the presented tool include short ($\sim$ 1 minute) processing time and improved segmentation quality. This is the first study to perform quick and accurate segmentation of 102 brain regions based on the surface-based protocol (DMK protocol), widely used by experts in the field. This is also the first work to include an expert reader study to assess the quality of the segmentation obtained using a deep-learning-based model. We show the superior performance of our deep-learning-based models over the traditional segmentation tool, Freesurfer. We refer to the proposed deep learning-based tool as DARTS (DenseUnet-based Automatic Rapid Tool for brain Segmentation). Our tool and trained models are available at \url{https://github.com/NYUMedML/DARTS}
\end{abstract}


\section{Introduction}
Quantitative regional brain volumetrics have been used to study nearly every neurological, developmental and behavioral condition known, from normal aging\cite{peters2006ageing,cole2015prediction,pini2016brain}, to schizophrenia\cite{van2016subcortical} to dementia\cite{bron2015standardized, coupe2019lifespan, moscoso2019prediction,devanand2008combining, devanand2007hippocampal}, to multiple sclerosis\cite{de2014clinical, azevedo2019contribution} and hydrocephalus \cite{nph}, just to name a few. Semi-automated segmentation tools have been widely applied for this task, providing measures of whole-brain and regional brain volumes \cite{freesurfer},\cite{FSL}. Segmentation is useful for multiple other research tasks such as coregistration with other imaging modalities (e.g., functional MRI, diffusion MRI, Positron emission tomography) and anatomical localization. Influential research initiatives including the Human Connectome Project \cite{van2013wu}, Alzheimer's Dementia Neuroimaging Initiative (ADNI) \cite{adni}, National Alzheimer's Coordinating Center (NACC) \cite{nacc}, and the UKBioBank \cite{UKbiobank} rely on traditional brain segmentation tools to provide quantitative measures to researchers.

Traditional semi-automated methods are based on Markov random fields and apply boundary determination methods to separate cortical gray matter from subcortical white matter. Despite such tools 1) being available through both open source \cite{freesurfer} and commercial visualization products for decades, and 2) having clear potential utility, this technology has failed to translate well to routine clinical care, in part due to the need for manual corrections and off-line processing that can take hours even with modern computing capabilities. In the clinical setting, these aspects place significant practical barriers to successful implementation. 

Recent innovations using deep learning for solving problems in computer vision have resulted in a revolution in medical imaging \cite{survey}. In particular, there have been novel developments using deep learning for medical imaging segmentation tasks \cite{unet,vnet,refinenet,deeplab}.

Previous efforts applying deep-learning-based models to a brain segmentation task \cite{voxnet_3_structures,quicknat,deepnat,patch_based_3_structures,3d_dense_3_structures,3d_conv_8_structures} provide proof of concept that segmentation for coarse regions of interest (ROIs) ($\sim$ up to 35 regions) is promising. The major practical limitation of these prior works is incomplete segmentation of the brain into finer anatomic regions which are typically available through traditional tools like Freesurfer. There are substantial challenges in terms of how to approach the segmentation of these finer anatomic regions, relating to the small size of these regions containing relatively few voxels and the resulting class imbalance. 

Here, we extend the use of deep-learning-based models to perform segmentation of a complete set of cortical and subcortical gray matter structures and ventricular ROIs, matching the regions included in the commonly used, standard tool, Freesurfer (aseg+aparc segmentation libraries), to provide a real-life rapid brain segmentation tool. We employ a weighted loss function, weighing each ROI in inverse proportionality to its average size to address the extreme class imbalance. Additionally, we use dense convolutions in the U-net architecture and show that such architecture (called DenseUNet) provides us substantial gains over the baseline U-net model in terms of Dice Score improvement.  

We assess both the quality of segmentation obtained using our deep-learning-based model and time required for segmentation compared against standard Freesurfer segmentation using both quantitative indices as well as expert evaluation. To our knowledge, this is the first report with accompanying source code of a practical tool that can be used both in a research setting to augment standard prior methods as well as in clinical settings to provide fast and accurate quantitative brain measures.

\section{Related Work}

\subsection{Current Tools for Brain Segmentation}
Many different brain segmentation tools such as Freesurfer \cite{freesurfer}, STAPLE \cite{staple} and PICSL \cite{wang2013multi} are currently used by neuroimaging researchers and radiologists. All of these tools are based on atlas registration via nonrigid registration methods, which are computationally expensive during inference. Of the tools mentioned above, Freesurfer is one of the most commonly used tools. Freesurfer is based on topological surface mappings to detect gray/white matter boundaries followed by nonlinear atlas registration and nonlinear spherical surface registration for each sub-cortical segment. Each step involves an iterative algorithm, and the surface registration is based on inference on Markov Random Fields (MRF) initially trained over manually labeled datasets\cite{freesurfer, reuter2012within}. Despite the surface registration method being spatially non-stationary, due to 1) non-convex nature of the model, 2) subject- and pathology-specific histological factors that impacts intensity normalization, and 3) iterative process for finding optimal segmentation, Freesurfer creates different results under different initialization settings, even for the same scan. It is known that Freesurfer outputs different results if the previous scans of the patient are taken into account \cite{reuter2012within}. First released in 2006, Freesurfer has been used innumerable times by researchers, saving the need to perform complete manual segmentation of brain MRIs, which was the prior standard; however, the methodology employed by this tool and others like it suffer from some inherent limitations. Specifically, each transformation of Freesurfer on a single brain volume is computationally intensive and the time required to segment a single 3D MRI volume can be on the order of hours. Additionally, the quality of segmentation of such MRF-based models is also lower than the deep-learning-based models which are demonstrated in the reader study performed in this report. Similar limitations plague all of the other traditional tools.

\subsection{Deep-learning-based Brain Segmentation Tools/Models}
With the advent of deep learning methods for computer vision tasks like classification, object detection, and semantic segmentation, some of the inherent limitations of traditional image processing methods were resolved \cite{lecun2015deep}. Consequently, these techniques were employed in several application domains including segmentation of brain cortical structures. 
Researchers have approached the task of segmentation of brain both by using 2D slices \cite{2d_paper_1,2d_paper_2} and 3D volumes as inputs \cite{deepnat,slant,3d_dense_3_structures,3d_conv_8_structures}. Despite 3D models naturally utilizing 3D structural information inherent in brain anatomy, it has been shown that such models do not necessarily yield superior results \cite{3d_not_superior, 2d_paper_2}. Additionally, they tend to be computationally more expensive and therefore slower during inference. 3D-based whole volume methods also require a pre-defined number of slices through the brain volume as input and, in practice, the number of slices varies between protocols, making such models potentially less generalizable. Researchers including \cite{slant}, and \cite{deepnat} have attempted to address the computational cost by training on volumetric patches; however, inference time remains relatively long (DeepNAT requires $\sim$1-2 hours and SLANT takes $\sim$15 mins using multiple GPUs).  \cite{2d_paper_1} and \cite{2d_paper_2} have performed segmentation using patches from a 2D slice through the brain, offering 3 and 10 segments respectively. But compared with the over 100 segments available via Freesurfer, these few segments limit the tools' practical utility. 
In order to take advantage of 3D information while keeping the computational cost low, in QuickNat \cite{quicknat}, 2D convolutional neural networks in multiple planes have been trained and combined, but this also requires a complete 3D volume with voxel resolution being $1mm^3$. To perform such a preprocessing, Freesurfer is needed. Additionally, QuickNat only provides coarse segmentation for $\sim$ 30 ROIs making it less usable for clinical purposes.  


These prior works like \cite{slant} and \cite{deepnat} clearly show the promise that deep learning models can be used for segmenting the brain into anatomic regions; however, in some prior models, the potential benefit in computation derived from using a deep-learning-based approach is negated by the need for slow pre-processing steps (e.g. registration, intensity normalization, conforming in \cite{slant} ) or post-processing steps (e.g., Conditional Random Field in \cite{deepnat}) that are required for these tools to operate. These steps increase the computational cost of the complete pipeline and render them slower.

In summary, our goal is to provide a tool with high accuracy, short inference time and sufficient brain segments to be useful in current research practice and clinical applications.   

Contributions of this work are as follows:
\begin{itemize}
    \item To the best of our knowledge, this is the first presentation of a truly practical, deep-learning-based brain segmentation tool that can provide accurate segmentation of over 100 brain structures, matching regions found in aseg+aparc segmentation libraries from one of the leading industry-standard, registration-based tools, Freesurfer. 
    \item Here, we leverage the benefits of using a surface-based approach (specifically DMK protocol) for brain segmentation.
    \item We impose no additional, registration-based pre-processing or post-processing steps and achieve inference times of $\sim$1 minute using just a single GPU machine.
    \item We show an excellent generalization of our model to different MRI scanners and MRI volumetric acquisition protocols.
    \item In addition to quantitative assessments against Freesurfer segmented data, we also evaluate our model against manually segmented data and perform an expert reader study to assess the quality of our segmentation tool.
\end{itemize}














\section{Methods}

The study is conducted in compliance with the local Institutional Review Board at NYU Langone Health.

\subsection{Data} \label{sec:data_desc}
The training data comes from the Human Connectome Project (HCP) \cite{HCP}. Specifically, we used 3D image volumes from a Magnetization Prepared Rapid Acquisition Gradient Echo (MPRAGE) volumetric sequence obtained as part of the HCP protocol. These images were acquired at multiple sites at 3 Tesla (Connectome Skyra scanners; Siemens, Erlangan) with the following parameters: FOV = 224mm x 224mm, resolution = 0.7mm isotropic, TR/TE = 2400/2.14 ms, bandwidth = 210 Hz/pixel. Each MRI volume has a dimension of 256 $\times$ 256 $\times$ 256. For training the model, a 2-D coronal slice of the MRI volume was used as the input. The coronal plane was selected based on superior performance in our preliminary experiments compared to axial and sagittal. The box-plots for the dice scores of these experiments can be seen in the appendix section (Figures \ref{fig:compare_dice_plot_aparc_hcp_cva_part_1}, \ref{fig:compare_dice_plot_aparc_hcp_cva_part_2}, \ref{fig:compare_dice_plot_aparc_hcp_cvs_part_1}, and \ref{fig:compare_dice_plot_aparc_hcp_cvs_part_2}).  Freesurfer segmentation results processed for the HCP (\cite{HCP}) were used as auxiliary ground truth for initial training of the models. These segmentation results had undergone manual inspection for quality control \cite{marcus2013human}. Further, to fine-tune the model, we used two additional data sources: manually segmented data by \cite{DKT} and Freesurfer segmented, manually corrected NYU data.

Initially, we focused on 112 regions per individual 3D brain volume based on Freesurfer (aseg+aparc) labels. Figure \ref{fig:voxel_count_top} and figure \ref{fig:voxel_count_bottom} show the voxel distribution of the 112 segments in our study, revealing a class imbalance challenge: only a few regions are large (>60000 voxels) whereas most of the regions are significantly smaller (<20000 voxels). The class imbalance challenge is addressed in section \ref{sec:loss_func}. 
The following 10 regions were excluded from the analysis: segments labeled left and right 'unknown', four brain regions not common to a normal brain: White matter and non-white matter hypointensities and left and right frontal and temporal poles, and segments without widely accepted definitions in the neuroradiology community (left and right bankssts) \cite{DKT}. 

We randomly divided the cohort into training, validation and (held out) test sets with 60\% (667 scans), 20\% (222 scans) and 20\% (222 scans) ratio. We separated training, validation and held-out test sets according to patients rather than slices to prevent data leakage.  

\subsection{Manually Annotated Dataset}
We used Mindboggle-101 \cite{DKT} manually annotated data to fine-tune and further evaluate our model. The dataset includes manual annotations for all the segments that are present in the Freesurfer aseg+aparc list (except those listed in the previous section). 
The Mindboggle-101 dataset contains MRI scans of normal individuals from a diverse number and type of MRI scanners ranging in magnetic field strength from 1.5 Tesla to 7 Tesla. Mindboggle-101 contains 101 MRI scans from 101 subjects from multiple open-source studies. The details of the MRI acquisition for each dataset can be found in table \ref{table:mindboggle_mri_aq}. These data were also randomly split into training, validation and (held out) test set with the same 60\%(60 scans), 20\%(21 scans) and 20\%(20 scans) ratio, again separated according to patients rather slices to prevent data leakage. The subjects' ages range from 20 to 61 years. 

\begin{table}[]
    \centering
    \begin{tabular}{p{30mm}|p{15mm}|p{40mm}|p{55mm}}
         Dataset Name & Num. of Subjects & Scanners & Acquisition Details  \\
         \hline
         \hline
         Oasis-TRT \cite{oasis} & 20 & MP-RAGE 1.5-T  Vision  scanner (Siemens,  Erlangen,  Germany) & Resolution = 1mm $\times$ 1mm $\times$ 1mm, TR/TE = 9.7/4 ms \\
         \hline
         Multi-modal MRI Reproducibility \cite{mmr}  & 21 & MP-RAGE 3T MR scanner (Achieva, Philips Healthcare, Best, The Netherlands) & Resolution = 1mm $\times$ 1mm $\times$ 1.2mm, TR/TE/TI =6.7/3.1/842ms, FOV =  240 $\times$ 204 $\times$256mm \\
         \hline
         Multi-modal MRI Reproducibility 3T/7T \cite{mmr}  & 2 & MP-RAGE 3T/7T MR scanner (Achieva, Philips Healthcare, Best, The Netherlands) & Resolution =  1mm $\times$ 1mm $\times$ 1.2mm, TR/TE/TI =6.7/3.1/842ms, FOV =  240 $\times$ 204 $\times$256mm \\
         \hline
         Nathan Kline Institute/Rockland Sample  & 22 & MP-RAGE MR  Siemens Magnetom & Resolution = 1mm $\times$ 1mm $\times$ 1mm, TR/TE =1900/2.52, FOV =  250 $\times$ 250 $\times$250mm, Bandwidht = 170 Hz/Px \\
         \hline
         Nathan Kline Institute/Test-Retest  & 20 & MP-RAGE MR Siemens Magnetom & Resolution = 1mm $\times$ 1mm $\times$ 1mm, TR/TE =1900/2.52, FOV =  250 $\times$ 250 $\times$250mm, Bandwidht = 170 Hz/Px \\
         \hline
         Human Language Networks \cite{HLN} & 12 & MP-RAGE 3T MRI scanner (Philips Medical Systems, Best, Netherlands) & FOV = 240 mm, TE = 35 ms, TR = 2 sec, 4.5 mm thickness \\
         \hline
         Colin Holmes Template \cite{holmes}  & 1 & MP-RAGE Phillips 1.5 T MR & Resolution = 1mm $\times$ 1mm $\times$ 1mm, TR/TE = 18/10 ms, FOV = 256 mm (SI) $\times$ 204 mm (AP) \\
         \hline
         Twins-2 \cite{DKT}  & 2 & MP-RAGE &  \\
         \hline
         Afterthought-1 \cite{DKT} & 1 & MP-RAGE & \\
         \hline
    \end{tabular}
    \caption{MRI acquisition details of all the data in Mindboggle-101 dataset}
    \label{table:mindboggle_mri_aq}
\end{table}{}

\subsection{NYU's Manually Corrected Dataset}
We also use a small internal NYU dataset consisting of 11 patients to train and assess the generalizability of the segmentation model. The description of the NYU dataset is as follows: MPRAGE (FOV=256 $\times$ 256 mm$^2$; resolution=1 $\times$ 1 $\times$ 1 mm$^3$; matrix=256 $\times$ 256; sections, 192; TR=2100 ms;TE=3.19 ms; TI=900 ms; bandwidth=260 Hz/pixel). Imaging was performed on 3T Siemens Skyra and Prisma MRI scanner. Each MRI scan in this dataset was first processed using the standard Freesurfer tool and then underwent manual corrections by an expert neuroimager for ground truth segmentation. Here also, we split the data into the train (6 scans), validation (2 scans) and held-out test (3 scans) sets.

\subsection{Data Augmentation}
Differences across MRI scanners and acquisition parameters result in differences in image characteristics such as signal-to-noise, contrast, the sharpness of the acquired volume. In addition, there are between-subject differences arising from subject positioning, etc. 
To improve the generalizability of our models to a broad array of scanner parameters, one or more of the following augmentation methods was applied to the training data at a random 50\% of the training batches: First, gaussian blurring (sigma parameter uniformly drawn from [0.65 to 1.0]) and gamma adjustment (gamma parameter uniformly drawn from [1.6 to 2.4]).
Second, input MRI and corresponding target labels were rotated by angle theta where theta is a random number uniformly drawn from [-10 to +10] degrees. Third, input and the corresponding labels were shifted up and sideways by dx and dy pixels, where dx and dy were randomly and uniformly drawn between +25 and -25. The input was then normalized between 0 and 1 using min-max normalization. Data augmentation was used only during training.

\subsection{Deep Learning Model}
\begin{figure}[t]
    \centering
    \includegraphics[width=\linewidth]{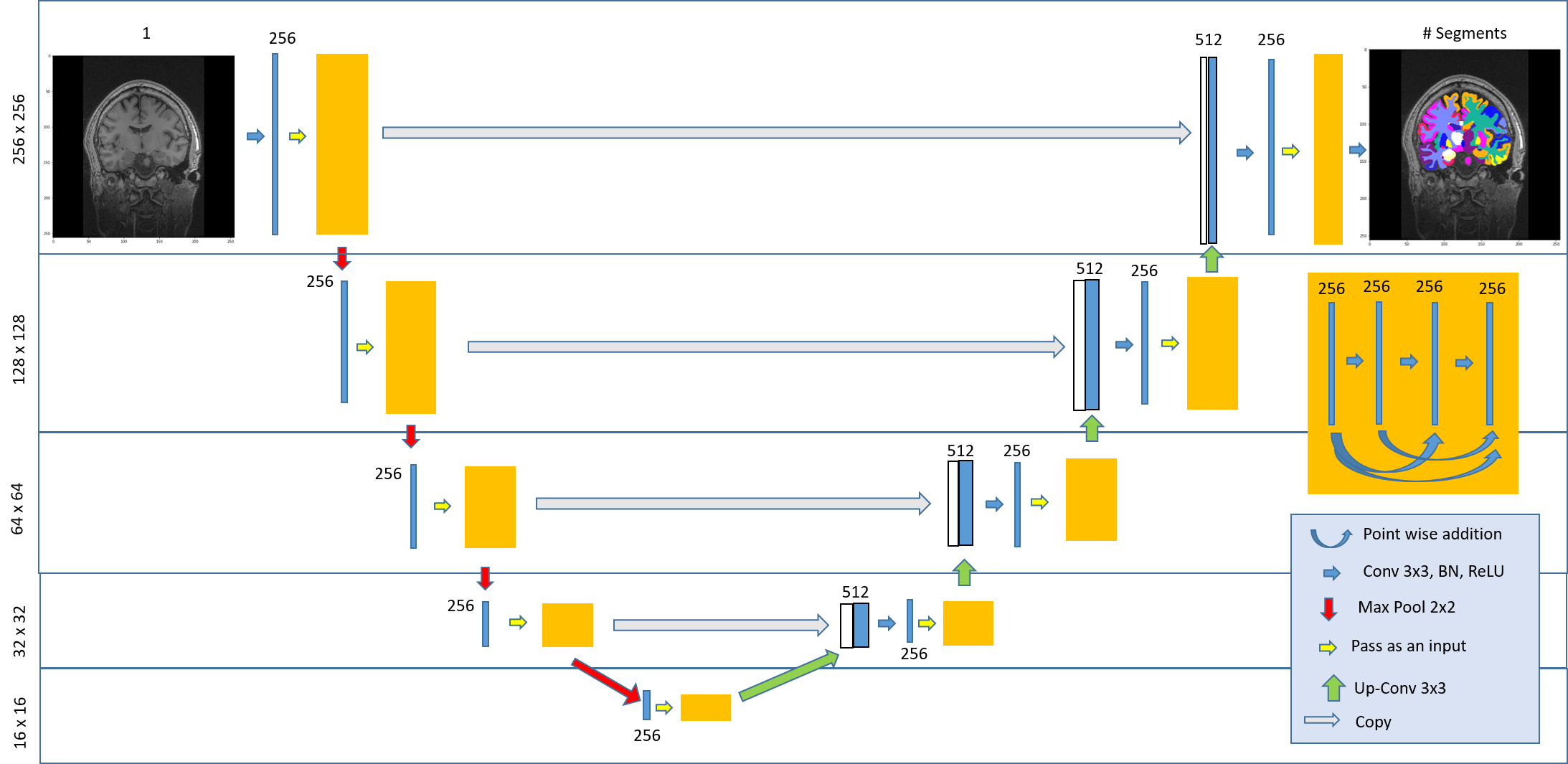}
    \caption{Schematic Diagram of DenseUNet}
    \label{fig:DenseUnet}
\end{figure}

The model architecture used for performing the segmentation task is an encoder-decoder style fully convolutional neural network. The architecture is partly inspired by U-Net architecture \cite{unet} and partly by \cite{tiramisu} and \cite{quicknat}. We term the new architecture as DenseUNet. The DenseUNet has a U-Net like architecture where it has four dense blocks in the encoding path and four dense blocks in the decoding path. The encoding pathway is connected to the decoding pathway using another dense block called the connecting dense block. Each of the encoding dense blocks is connected to the corresponding decoding dense block using skip connections. These skip connections facilitate the gradient flow in the network as well as the transfer of spatial information required for decoding the encoded input. The output from the last decoding dense block is passed through a classification block where the network performs the classification and gives the final output in the form of a separate probability map for each brain structure. The schematic diagram of the entire architecture can be seen in figure \ref{fig:DenseUnet}.

We also implemented a vanilla U-Net architecture as described in the original paper \cite{unet} which serves as a baseline model. The schematic diagram for the same can be seen in figure \ref{fig:unet}.

The architectural choice of DenseUNet was also motivated by our empirical results on U-Net. A Dense block has more model capacity compared to standard convolutional block \cite{densenets}. When larger training data is available, DenseUNet has larger learning and generalization capabilities.

All the parameters of the U-net and DenseUNet were initialized using Xavier initialization \cite{xavier}.

The components of DenseUNet i.e. the encoding dense block, the connecting dense block, the decoding dense block, and the classification block are explained below.

\subsubsection{Encoding Dense Block}
The encoding dense block has four convolution layers. The output of all the previous convolution layers is added to the subsequent convolution layers. These additive connections are termed dense connections, that facilitate the gradient flow and allow the network to learn a better representation of the image \cite{densenets}. The other common connection type is a concatenated connection. QuickNAT \cite{quicknat} uses concatenate connections to build dense blocks. Though concatenated connections can model additive connections, the model complexity in terms of the number of parameters and number of mathematical operations increases significantly leading to out-of-memory issues while training the model for a large number of segments. Therefore, to avoid out-of-memory issues and to achieve low training and inference times, we choose to use additive dense connections as opposed to concatenating dense connections. The output obtained by adding all the previous convolution layers' output is followed by batch normalization and Rectifier Linear Unit (ReLU) non-linearity. Each convolution layer has 256 output channels with the filter size being 3 $\times$ 3 for all the channels. The output of the encoding dense block is followed by a max-pooling layer with a kernel size of 2 $\times$ 2 and a stride of 2. The down-sampled output is fed to the next encoding or connecting dense block.

\subsubsection{Connecting Dense Block}
The connecting dense block is similar to the encoding dense block and has four convolution layers with dense connections. The only difference is the output of the dense block is not followed by a downsampling layer like a max-pooling layer.
\subsubsection{Decoding Dense Block}
The decoding dense block is preceded by an upsampling block. The output from the previous decoding or connecting dense block is upsampled using transposed convolution with a filter size of 4 $\times$ 4 and stride of 2 and padding of 1. The upsampled output is concatenated with the output from the corresponding encoding dense block. The concatenated output serves as an input to a convolution layer which is followed by batch normalization and ReLU activation. The convolution layer has 256 output channels with a filter size of 3 $\times$ 3. 
\subsubsection{Classification Block}
The classification block is a single convolution layer with the number of output channels equal to the number of brain structures to segment (in our case 112 \footnote[1]{\label{note1}Our architecture trains for full aparc+aseg segments i.e. 112 segments, and we omit the 10 excluded segments mentioned in section \ref{sec:data_desc} after training.}) with a filter size of 1 $\times$ 1. The output of the convolution layer is passed through a softmax layer to obtain the probability maps for each of the brain structures we are trying to segment.

\subsection{Loss Function} \label{sec:loss_func}
We model the segmentation task as a multi-class classification problem. Here, since we have $112$
\footnoteref{note1}
tissues of interest, this is a $113$-class classification problem, where the last class is "background". 

Since the dataset is an imbalanced one, we use a weighted cross-entropy loss and weighted dice loss for our task.  
The weighted cross entropy loss is defined as:
\begin{equation}
\text{Weighted-CEL} = -\frac{1}{N}\sum_{i=1}^N{\sum_{j=1}^{S}}{w_j(y_{ij}log(p_{ij}))}
\end{equation}
Where $w_j$ = Weight of the jth segment and S is the total number of segmentation classes. Here, $w_j$ = $\frac{\text{median freq}}{\text{freq(j)}}$, where, $\text{freq(j)}$ is the number of pixels of class $j$ divided by the total number of pixels of images where $j$ is present, and $\text{median freq}$ is the median of these frequencies \cite{fergus_median_freq}.  N = number of pixels in a 2D MRI image (a slice of the MRI volume), $p_{ij} = $ probability of pixel i to be belonging to segment j, $y_{ij} = $ label of pixel i to be belonging to segment j = 1 or 0.)

Weighted Dice Loss: The weighted dice score is defined as:
\begin{equation}
\text{Weighted-Dice Loss} = 1-2\frac{\sum_{j=1}^{{S}}w_j\sum_{i=1}^Ny_{ij}p_{ij}}{\sum_{j=1}^{{S}}w_j\sum_{i=1}^Ny_{ij}+p_{ij}}
\end{equation}
Weights are calculated using the same approach as mentioned for weighted-CEL. 

We experiment with local(using only a specific batch) and global (using the entire train set) estimation of $w_j$. In the local case,  $w_j$s were adapted for each batch, and hence the loss function for each batch was slightly different. However, we found that a global $w_j$ gave us the best out of sample performance.

We combine the above two losses in a novel way that we term as loss switching. Loss switching is explained in the next section.

\subsection{Loss Switching} \label{sec:loss_sche}
In segmentation tasks, the dice score is often reported as the performance metric. A loss function that directly correlates with the dice score is the weighted dice loss \cite{wt_dice_loss}. Based on our empirical observation, the network trained with only weighted dice loss was unable to escape local optimum and did not converge. Also, empirically it was seen that the stability of the model, in terms of convergence, decreased as the number of classes and class imbalance increased. We found that weighted cross-entropy loss, on the other hand, did not get stuck in any local optima and learned reasonably good segmentations. As the model's performance with regard to dice score flattened out, we switched from weighted cross entropy to weighted dice loss, after which the model's performance further increased by 3-4 \% in terms of average dice score. This loss switching mechanism, therefore, is found to be useful to further improve the performance of the model.



\subsection{Evaluation Metric}
Dice score (DSC) is employed here as a primary measure of quality of the segmented images. This is a standard measure against which segmentations are judged and provide direct information on similarity against the ground truth labels. 
\begin{equation}
\text{DSC} = \frac{2||PT||_2^2}{||P||_2^2+||T||_2^2}
\end{equation}

where P = Predicted Binary Mask, T = True Binary Mask, PT = element-wise product of P and T, $||X||_2$ is the L-2 norm. Dice score can equivalently be interpreted as the ratio of the cardinality of ($T \cap P$) with the cardinality of (($T \cup P$) + ($T \cap P$)).

From the above definition, it can be seen that DSC penalizes both over prediction and under prediction and, hence, is well-suited for segmentation tasks such as the one proposed here and is particularly useful in medical imaging.

\subsection{Training Procedure}
What follows is a description of the training procedure we employed:
\begin{itemize}
    \item For training each model, we used Adam Optimizer with reducing the learning rate by a factor of 2 after every 10-15 epochs.
    \item The model was initially trained on the training set of scans from the HCP dataset with the auxiliary labels until convergence. The trained model was then finetuned using the training set of the manually annotated dataset (Mindboggle-101) and the training set of the in-house NYU dataset. 
    \item We trained the model with the HCP dataset using the loss switching procedure described in section \ref{sec:loss_sche} whereas, for finetuning, the loss function is simply the weighted dice loss as described in section \ref{sec:loss_func}.
    \item All the models are trained using early stopping based on the best dice score on the validation set.
\end{itemize}

\section{Methodology for Reader Study} \label{sec:reader_study}
\subsection{Reader Study: Description and Setup}
We perform an expert reader evaluation to measure and compare the deep learning models' performance with the Freesurfer model. We use HCP held-out test set scans for reader study. On these scans, Freesurfer results have undergone a manual quality control\cite{marcus2013human}. We also compare the non-finetuned and fine-tuned model with the Freesurfer model with manual QC. Seven regions of interest (ROIs) were selected
: L/R Putamen (axial view), L/R Pallidum (axial view), L/R Caudate (axial view), L/R Thalamus (axial view), L/R Lateral Ventricles (axial view), L/R Insula (axial view) and L/R Cingulate Gyrus (sagittal view). The basal ganglia and thalamus were selected due to their highly interconnected nature with the remainder of the brain, their involvement in many neurological pathologies, and their ease of distinction from surrounding structures. The insular and cingulate gyri were selected to assess the quality of cortical segmentation in structures best visualized in different planes and also due to the relatively frequent involvement of the insular gyrus in the middle cerebral artery infarctions. The lateral ventricles were selected to assess for quality of segmentation of cerebrospinal fluid structures, which would help identify pathologies affecting cerebrospinal fluid volumes, including intracranial hypotension, hydrocephalus, and cerebral atrophy. 

\begin{figure}[t]
    \centering
    \includegraphics[width=\linewidth]{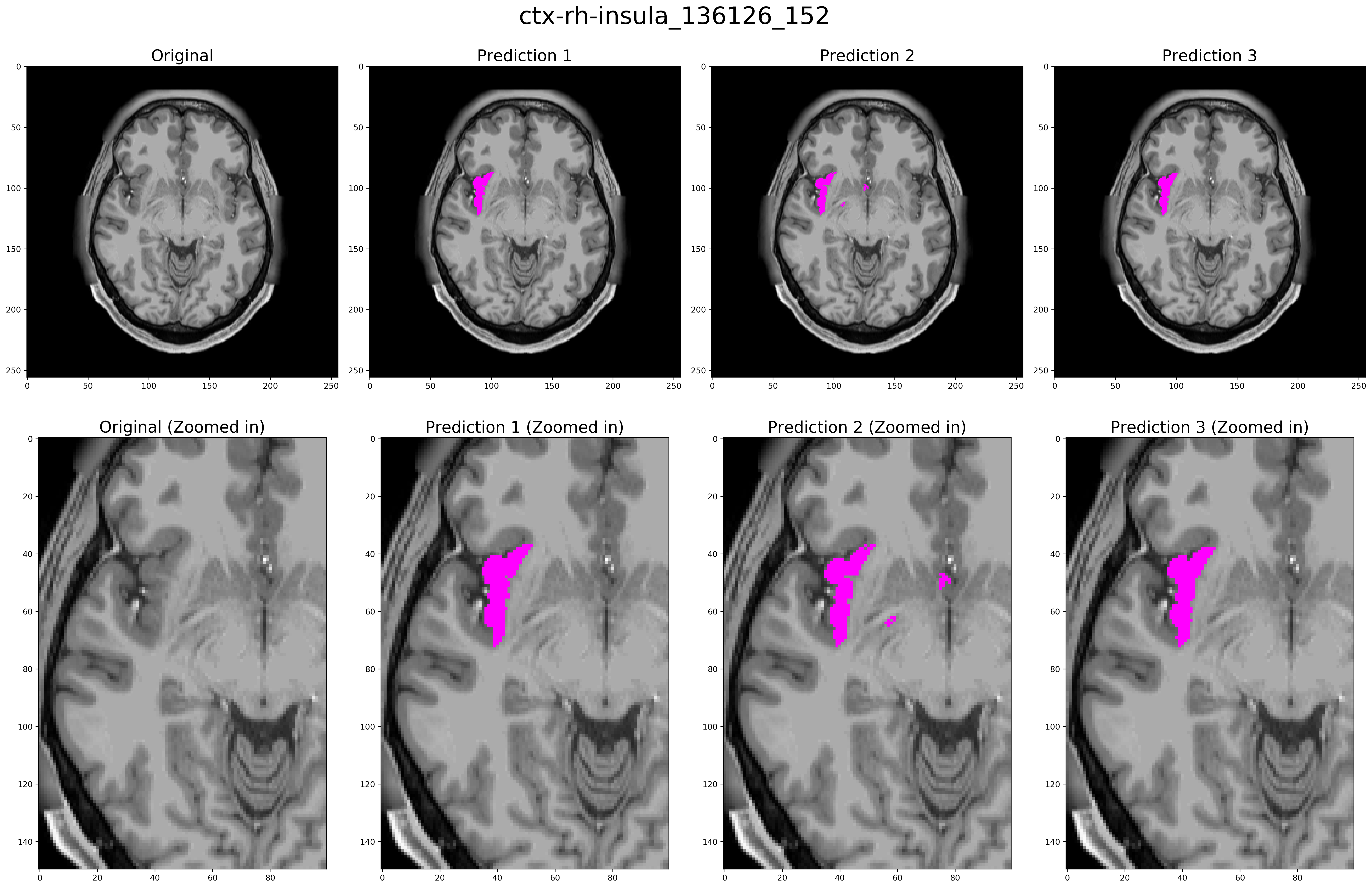}
    \caption{A sample segmentation of Right Insula used in the expert reader evaluation. Here, predictions 1, 2 and 3 are from the Finetuned model, Freesurfer, and non-Finetuned model respectively; though in the reader study,  numbering and order of presentation of the predictions are randomized and a total of 280 of examples are presented. Each reader is asked to rate each example for the quality of segmentation on a 5-point Likert-type scale.}
    \label{fig:sample_reader_study}
\end{figure}

Three expert readers performed visual inspection and assessment of the segmentation results. There were two attending neuroradiologists with 3 and 5 years of experience and one second-year radiology resident. Each reader was asked to rate 40 different predictions for each ROI (20 in each brain hemisphere) such as shown in figure \ref{fig:sample_reader_study}. Readers were blinded to the algorithm used to predict the segmentation and examples were presented in a randomized order. Each prediction presented consisted of a single slice containing a minimum of 40 pixels within the ROI, ensuring that enough of the structure being assessed was present on the given image. A sample slice rated by the readers is shown in figure \ref{fig:sample_reader_study}.

Each reader rated each example on a Likert-type scale from 1 to 5 with the following definitions:
\begin{enumerate}
    \item Rating of 1 (Poor): Segmentation has major error(s) in either the major boundary or labeling of areas outside the area of interest. Such segmentation would not yield acceptable quantitative measures.
    \item Rating of 2 (Fair): Segmentation has >2 areas of error along the major boundary or includes small but nontrivial areas outside the area of interest that would require manual correction before yielding reasonable quantitative measures to be used in research or clinical scenarios.
    \item Rating of 3 (Good): Segmentation has 1 or 2 area(s) of error along the major boundary that would require manual correction before yielding reasonable quantitative measures to be used in research or clinical scenarios. A good segmentation could have minor/few voxels separate from the volume of interest.
    \item Rating of 4 (Very Good): Segmentation has minor, small areas of error along the major boundary that would still yield reasonable quantitative anatomic measures without manual corrections, appropriate for research or clinical use.
    \item Rating of 5 (Excellent): Segmentation is essentially ideal, has no (or only tiny inconsequential/questionable areas) where manual corrections might be used and would yield highly accurate quantitative anatomic measures, etc. Should have no erroneous areas outside the segment of interest.
\end{enumerate}{}
\subsection{Reader Study: Analysis}
Using the ratings obtained from three readers the following analyses are performed:
\begin{enumerate}
    \item Inter-Reader Reliability (IRR): An IRR analysis is performed using a two-way mixed, consistency, average measures ICC (Inter Class Correlation) \cite{mcgraw1996forming} to assess the degree of agreement between readers. High ICC indicates that the measurement error introduced due to independent readers is low and hence, the subsequent analysis' statistical power is not substantially reduced.
    \item Comparison of different models: Based on the readers' ratings, we investigate if there are statistically significant differences between the three methods using paired T-test and Wilcoxon signed-rank test at 95\% significance level.
\end{enumerate}{}

\section{Results}
Here, we report our quantitative evaluation results on the held-out test sets from manually annotated and corrected Mindboggle-101 and NYU dataset. 
We also report results of a qualitative evaluation via a reader study with expert neuroimaging radiologists on held out HCP scans and their corresponding Freesurfer labels and model's prediction.

\subsection{Quantitative Evaluation: Performance on the Manually annotated Dataset - Mindboggle-101}
Table \ref{table:mindboggle_results} includes the performance of the Finetuned model and the non-Finetuned model on the manually annotated test set from Mindboggle-101 data.

\begin{table}[h!]
\centering
\begin{tabular}{p{30mm}|p{30mm}|p{30mm}}
 Model Name & non - Finetuned & Finetuned  \\
 \hline
 \hline
 UNet (Baseline) & 0.7329$\pm$0.014 & 0.80$\pm$0.013  \\
 \hline
 DenseUNet  & 0.7431$\pm$0.015 & \bf{0.819$\pm$0.011} \\
 \hline
\end{tabular}
\caption{Mean Dice Score on 102 segments on Mindboggle-101 dataset. Here, non-Finetuned = Model trained using only HCP dataset, Finetuned =  Model initially trained on HCP dataset and subsequently finetuned using Mindboggle-101 and NYU Dataset}
\label{table:mindboggle_results}
\end{table}

\begin{figure}[ht]
    \centering
    \includegraphics[width=0.8\linewidth]{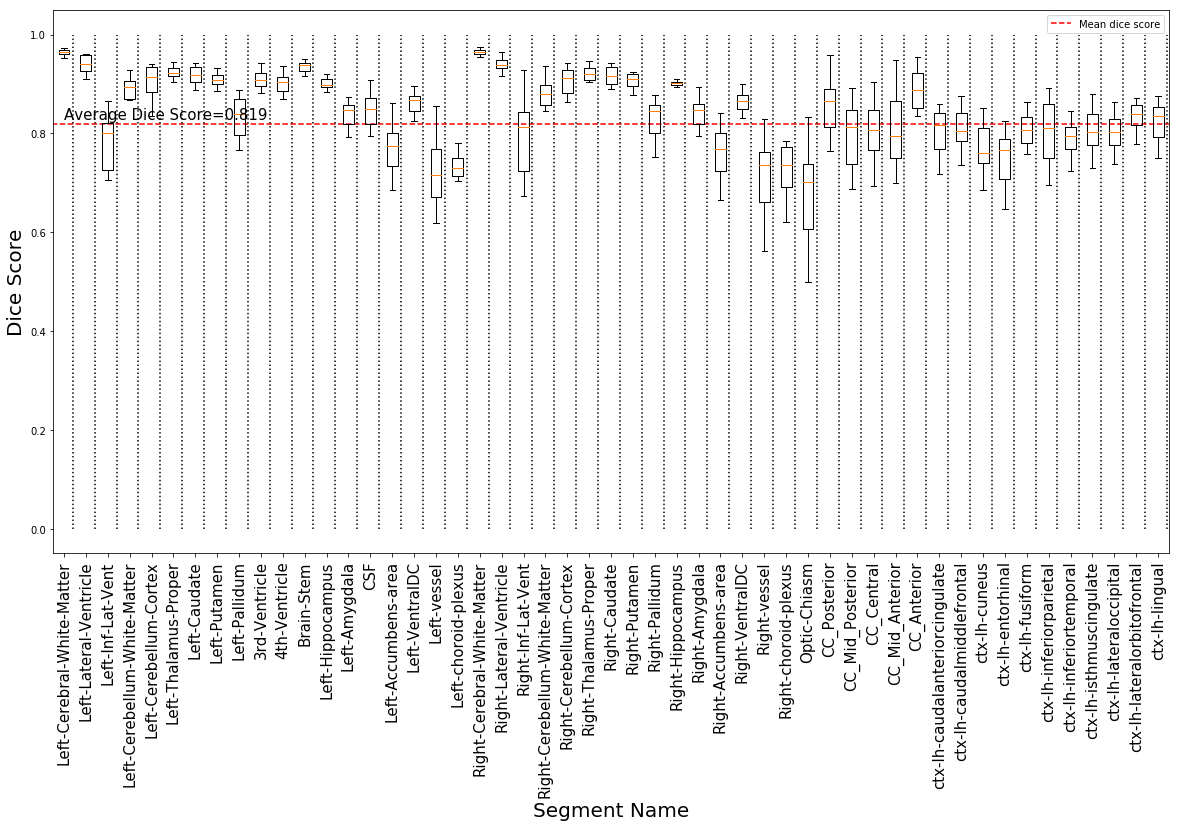}
    \caption{Box plot showing the dice scores for the first 51 regions of interest of Mindboggle dataset}
    \label{fig:dice_plot_man_part_1}
\end{figure}

\begin{figure}[ht]
    \centering
    \includegraphics[width=0.8\linewidth]{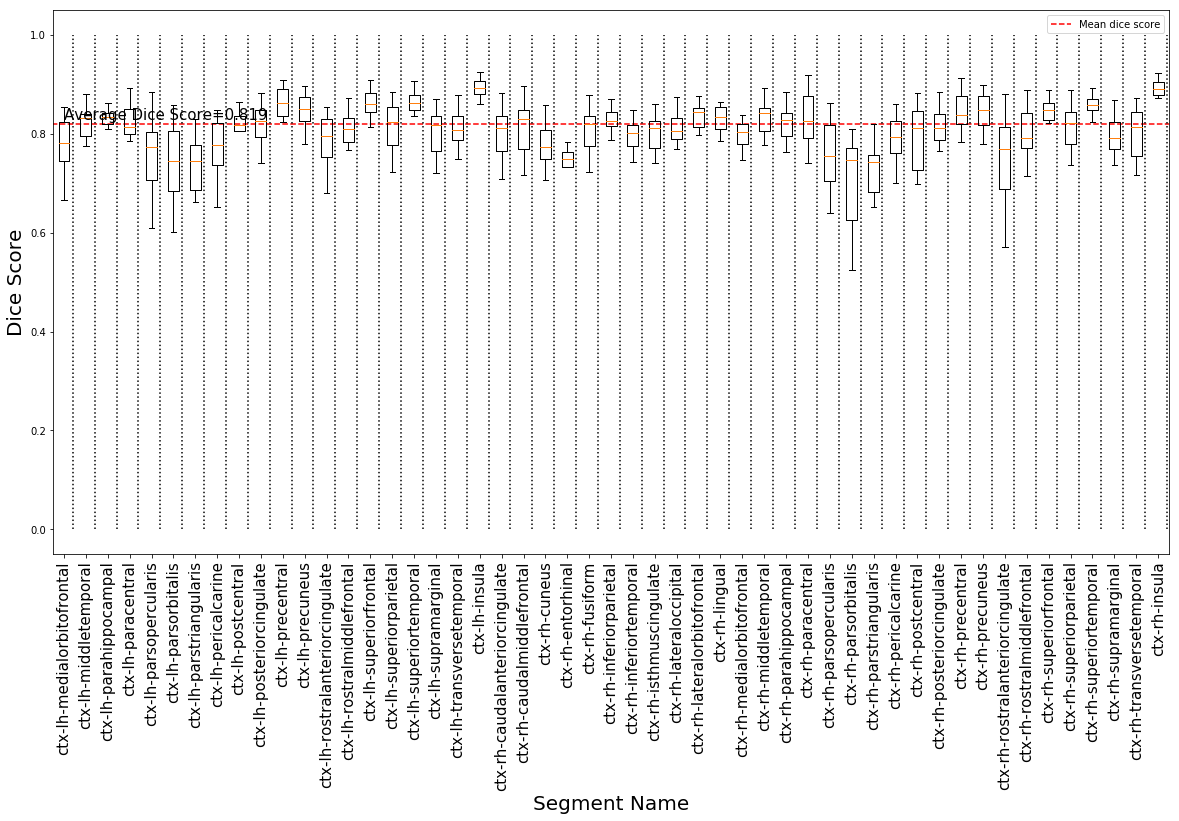}
    \caption{Box plot showing the dice scores for the next 51 regions of interest of Mindboggle dataset}
    \label{fig:dice_plot_man_part_2}
\end{figure}

Detailed dice scores for the 102 segments are included in figures \ref{fig:dice_plot_man_part_1} and \ref{fig:dice_plot_man_part_2}. More comprehensive comparison of fine-tuned vs non-fine-tuned results are also included in the Appendix (Figure \ref{fig:compare_dice_plot_man_part_1}, figure \ref{fig:compare_dice_plot_man_part_2}).

\subsection{Quantitative Evaluation: Performance on an External NYU Dataset}
Performance of the Finetuned model and non-Finetuned model on the external manually corrected NYU Test dataset is presented in the table \ref{table:nyu_results}.

\begin{table}[h!]
\centering
\begin{tabular}{p{30mm}|p{30mm}|p{30mm}}
 Model Name & non - Finetuned & Finetuned  \\
 \hline
 \hline
 UNet (Baseline) & 0.787$\pm$0.01 & 0.785$\pm$0.014  \\
 \hline
 DenseUNet  & 0.795$\pm$0.013 & \bf{0.800$\pm$0.012} \\
 \hline
\end{tabular}
\caption{Mean Dice Score on 102 segments on NYU-Dataset dataset. Here, non-Finetuned = Model trained using only HCP dataset, Finetuned =  Model initially trained on HCP dataset and subsequently finetuned using Mindboggle-101 and NYU Dataset}
\label{table:nyu_results}
\end{table}

For detailed results of each segment (with a box plot), please refer to the figure \ref{fig:dice_plot_nyu_part_1} and figure \ref{fig:dice_plot_nyu_part_2} in the Appendix.
\subsection{Quantitative Evaluation: Time}
Time to perform segmentation, and comparison with other models can be seen in the table \ref{table:time_results}. Our model gives the most informative segmentation (102 segments) in the least amount of time ($\sim$ 1 min). 
\begin{table}[h!]
\centering
\begin{tabular}{p{40mm}|p{30mm}|p{40mm}}
 Model Name & Num. of Segments & Time (for one brain Scan)  \\
 \hline
 \hline
 DenseUNet (our Model) & 102 & 65.6 secs ($\pm$ 0.9353 secs) (Single GPU Machine) \\
 \hline
 Unet (our Baseline Model)  & 102 & 54.2 secs ($\pm$ 0.7908 secs) \\
 \hline
 FreeSurfer \cite{quicknat} & $\sim$190 & $\sim$4 hrs \\
 \hline
 PISCL \cite{quicknat} & $\sim$190 & $\sim$30 hrs \\
 \hline
 DeepNAT \cite{quicknat} & 27 & $\sim$1 hr (on a Multi-GPU Machine) \\
 \hline
 QuickNAT \cite{quicknat} & 27 & $\sim$20 secs (on a Multi-GPU Machine) \\
 \hline
 SLANT \cite{slant} & 133 & $\sim$15 mins (on a Multi-GPU machine) \\
 \hline
\end{tabular}
\caption{Time to perform segmentation for a single MRI scan}
\label{table:time_results}
\end{table}

\subsection{Qualitative Evaluation: Reader study} \label{sec:reader_study_results}
\subsubsection{Inter Reader Reliability}
As mentioned in the section \ref{sec:reader_study}, IRR is assessed using a two-way mixed, consistency, average-measures ICC (Inter Class Correlation). The result can be seen in table \ref{table:irr_results}.
\begin{table}[h!]
\centering
\begin{tabular}{p{40mm}|p{70mm}}
 Model & Inter Class Correlation (Lower Bound - Upper Bound at alpha = 95\%) \\
 \hline
 \hline
 Freesurfer & 0.66 (0.58 - 0.72) \\
 \hline
 non-Finetuned model  & 0.82 (0.78 - 0.85)\\
 \hline
 Finetuned model & 0.80 (0.76 - 0.84)\\
 \hline
\end{tabular}
\caption{Inter Class Correlation between expert readers in assessment of quality of segmentations shows excellent agreement between readers for our model.}
\label{table:irr_results}
\end{table}

Since the resulting IRR is in good and excellent ranges \cite{cicchetti1994guidelines} for all the ratings, the statistical power and significance of the paired T-test and Wilcoxon signed-rank test to compare different model's performance would be reliable.

\subsubsection{Comparison of Different Models with Freesufer}
The table \ref{table:reader_study_results} showcases the average ratings obtained by each model (including Freesurfer). It also showcases whether the difference between any two model is statistically significant or not (at 95\% significance level) using paired T-test and Wilcoxon signed rank test.
\begin{table}[ht!]
\centering
\begin{tabular}{|c|l|l|l|l|}
\hline
\multirow{2}{*}{\textbf{Region of Interest}} & \multicolumn{3}{c|}{\textbf{Mean Rating $\pm$ One standard deviation}} & \multicolumn{1}{c|}{\multirow{2}{*}{\textbf{\begin{tabular}[c]{@{}c@{}}Statistically Significant Difference (s) \\ (as per paired T-test and Wilcoxon test)\end{tabular}}}} \\ \cline{2-4}
 & \multicolumn{1}{c|}{\textbf{FS}} & \multicolumn{1}{c|}{\textbf{NFT}} & \multicolumn{1}{c|}{\textbf{FT}} & \multicolumn{1}{c|}{} \\ \hline
Insula & \multicolumn{1}{c|}{\begin{tabular}[c]{@{}c@{}}L-3.13$\pm$0.88\\ R-2.85$\pm$1.00\end{tabular}} & \multicolumn{1}{c|}{\begin{tabular}[c]{@{}c@{}}L-3.90$\pm$1.01\\ R-3.48$\pm$1.26\end{tabular}} & \multicolumn{1}{c|}{\textbf{\begin{tabular}[c]{@{}c@{}}L-4.23$\pm$0.83\\ R-4.21$\pm$0.83\end{tabular}}} & L/R - FS and NFT, NFT and FT, FS and FT \\ \hline
Caudate & \begin{tabular}[c]{@{}l@{}}L- 4.26$\pm$0.79\\ R- 3.97$\pm$0.75\end{tabular} & \begin{tabular}[c]{@{}l@{}}\textbf{L- 4.46$\pm$0.70}\\ R- 4.17$\pm$0.76\end{tabular} & \begin{tabular}[c]{@{}l@{}}L- 4.45$\pm$0.75\\\textbf{R- 4.26$\pm$0.67}\end{tabular} & L/R - FS and NFT, FS and FT \\ \hline
Cingulate-Gyrus & \begin{tabular}[c]{@{}l@{}}L- 2.59$\pm$0.76\\ R- 2.72$\pm$0.74\end{tabular} & \textbf{\begin{tabular}[c]{@{}l@{}}L- 2.91$\pm$0.97\\ R- 3.0$\pm$0.88\end{tabular}} & \begin{tabular}[c]{@{}l@{}}L- 2.53$\pm$1.05\\ R- 2.49$\pm$0.89\end{tabular} & \begin{tabular}[c]{@{}l@{}}L - FS and NFT, NFT and FT\\ R - FS and FT, FS and NFT, NFT and FT\end{tabular} \\ \hline
Lateral-Ventricles & \begin{tabular}[c]{@{}l@{}}L- 4.14$\pm$0.83\\ R- 4.16$\pm$0.73\end{tabular} & \begin{tabular}[c]{@{}l@{}}\textbf{L- 4.46$\pm$0.66}\\ R- 4.39$\pm$0.73\end{tabular} & \begin{tabular}[c]{@{}l@{}}L- 4.44$\pm$0.73\\ \textbf{R- 4.41$\pm$0.72}\end{tabular} & \begin{tabular}[c]{@{}l@{}}L - FS and FT, FS and NFT\\ R - FS and FT, FS and NFT\end{tabular} \\ \hline
Pallidum & \begin{tabular}[c]{@{}l@{}}L- 3.20$\pm$0.80\\ R- 3.07$\pm$0.65\end{tabular} & \begin{tabular}[c]{@{}l@{}}L- 3.30$\pm$0.79\\ R- 3.72$\pm$0.78\end{tabular} & \textbf{\begin{tabular}[c]{@{}l@{}}L- 3.87$\pm$0.85\\ R- 3.93$\pm$0.78\end{tabular}} & \begin{tabular}[c]{@{}l@{}}L - NFT and FT, FS and FT\\ R - FS and NFT, NFT and FT, FS and FT\end{tabular} \\ \hline
Putamen & \begin{tabular}[c]{@{}l@{}}L- 3.22$\pm$2.14\\ R- 3.10$\pm$1.04\end{tabular} & \begin{tabular}[c]{@{}l@{}}L- 3.16$\pm$1.11\\ \textbf{R- 3.44$\pm$1.13}\end{tabular} & \begin{tabular}[c]{@{}l@{}}\textbf{L- 3.23$\pm$1.16}\\ R- 3.19$\pm$1.13\end{tabular} & \begin{tabular}[c]{@{}l@{}}L - No difference is statistically significant \\ (as per paired T-test)\\ L - FS and NFT, FS and FT \\ (as per Wilcoxon test)\\ R - FS and NFT, NFT and FT\end{tabular} \\ \hline
Thalamus & \begin{tabular}[c]{@{}l@{}}L- 3.40$\pm$0.75\\ R- 3.28$\pm$0.83\end{tabular} & \begin{tabular}[c]{@{}l@{}}\textbf{L- 4.0$\pm$0.88}\\ R- 3.96$\pm$0.82\end{tabular} & \begin{tabular}[c]{@{}l@{}}L- 3.95$\pm$0.94\\ \textbf{R- 4.02$\pm$0.87}\end{tabular} & \begin{tabular}[c]{@{}l@{}}L - FS and NFT, FS and FT\\ R - FS and NFT, FS and FT\end{tabular} \\ \hline
All Regions & \begin{tabular}[c]{@{}l@{}}3.41 $\pm$ 1.12\end{tabular} & \begin{tabular}[c]{@{}l@{}}3.78$\pm$1.04\end{tabular} & \begin{tabular}[c]{@{}l@{}}\textbf{3.86$\pm$1.07}\end{tabular} & \begin{tabular}[c]{@{}l@{}}FS and NFT, NFT and FT, FS and FT\end{tabular} \\ \hline
\end{tabular}
\caption{Reader study results comparing Freesurfer(FS), Non-fine-tuned(NFT) and Fine-tuned(FT) models' segmentation on a total of 20 evaluations per ROI, per Left(L) and Right(R) hemisphere, per model. Statistical tests are performed at 95\% significance level.}
\label{table:reader_study_results}
\end{table}

\section{Discussion}
\subsection{First fast and accurate segmentation for 102 regions of interest, consistent with aseg+aparc}
This work presents a deep-learning based model that performs extensive segmentation of the brain into 102 ROIs, matching the regions provided by the aseg+aparc segmentation libraries of the tool Freesurfer with inference time of $\sim$1 minute for a single 3D brain volume. Freesurfer is a widely used tool by neuroimaging researchers as well as the basis for many industry products, and thus the regions labelled under aseg+aparc are useful for current research and clinical applications.

In our reader study, quality of segmentations of Freesurfer as well as our developed models are evaluated by neuroradiologists. We find that the deep learning model trained on Freesurfer labels leads to significantly improved segmentation quality in 13 out of 14 regions assessed. We note that compared with MRF-based segmentation which forms the basis for Freesurfer, our deep learning model improves the quality of segmentation by producing smoother boundaries that follow the anatomic border more closely. 
Finetuning on manually annotated Mindboggle data further improves quality of segmentations for Insula and Pallidum ROIs. Interestingly, these are the areas where Freesurfer has the most reported boundary errors. 

Although the Mindboggle dataset is useful for improving performance for most of the ROIs, there are regions such as the Cingulate gyrus that are better segmented by the non-finetuned model. 

Based on our findings, optimal finetuning of the model would be to only finetune for regions that are not already well segmented, using the manually annotated dataset. This is part of our future explorations.

\subsection{Model Generalizability and Usability}
The model's generalizability was tested using a held out dataset from the Mindboggle dataset as well as a held out dataset from NYU. Our model achieves a mean dice score of 0.819 and 0.80 on the two datasets, respectively, showcasing good generalization capabilities on the unseen data. 

Additionally, if we see the detailed box plot shown in figure \ref{fig:dice_plot_man_part_1} and \ref{fig:dice_plot_man_part_2}, we notice that the dice scores for few regions such as Optic-Chiasm and Right-vessel are relatively lower than the dice scores for other regions. We investigated such regions and found that the regions with low dice scores are the regions with low mean voxel count (see figure \ref{fig:dice_count_reg_plot} in appendix). This is likely due to the fact that small errors in segmentation represents as a higher percentage of a small sized segment. Hence, overall dice scores for smaller regions tend to be lower than the dice scores for larger regions. 

The use of 2D images as inputs also adds to the practical usability of this model. Often in real-life clinical and research practices, incomplete 3D image volumes may be encountered. In such cases, performing segmentation using our tool remains straightforward, distinct from most currently available semi-automated tools and 3D deep-learning-based models that require entire 3D brain volumes with pre-specified number of slices as inputs.

Another advantage of the proposed the model is relatively short inference time. As seen in table \ref{table:time_results}, the model provides fairly comprehensive segmentation in $\sim$ 1 minute, comparing favorably against other available tools. This, combined with lack of dependency on pre-processing, makes our tool feasible on-demand in a clinical setting. Moreover, this work adheres to Freesurfer's region-naming convention, already in wide use and familiar to neuroimagers.



\subsubsection{Errors of Freesurfer Segmentation: Putamen, Insula and Pallidum} \label{sec:Freesurfer errors}
\begin{figure}[ht]
    \centering
    \includegraphics[width=0.6\linewidth]{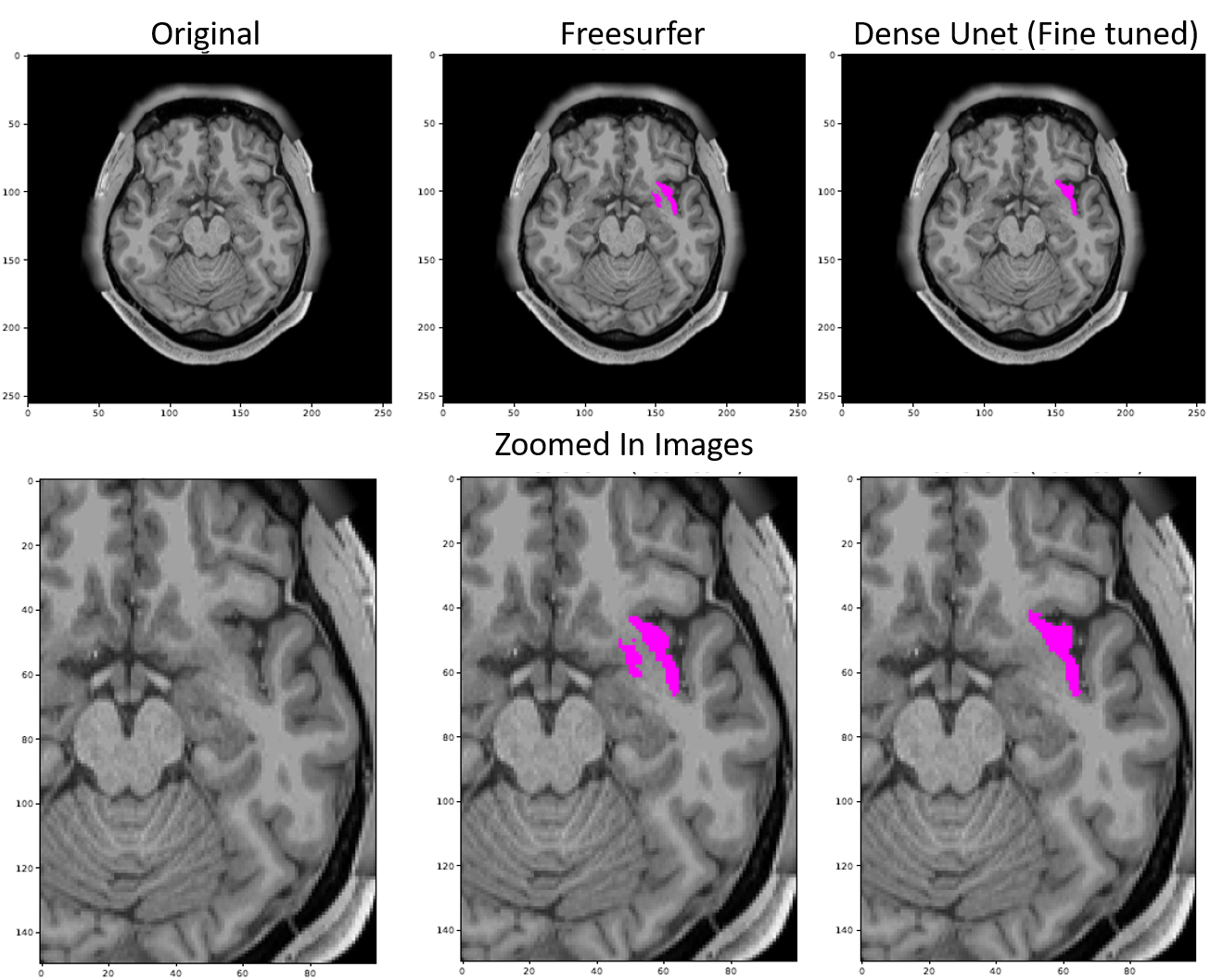}
    \caption{MR image in the axial plane through the level of the insula (left), Freesurfer (FS)  (center) and our proposed model, DenseUNet (right) prediction of the insula. It is evident that the FS prediction has errors in both over and underestimation along the boundary, includes discontiguous voxels and is also somewhat non-natural looking with stair-step artifact and a noisy appearance. DenseUNet insula segmentation, on the other hand, obeys well the segment boundaries and lacks the stair-step artifact, rendering a smooth contoured, more natural appearing segmentation. }
    \label{fig:faulty_seg_insula}
\end{figure}

The work of \cite{quality} reveals errors in Freesurfer segmentation, thus necessitating manual corrections for quality control. Our reader study results also demonstrates the low quality nature of Freesurfer segmentations and confirms the findings of \cite{quality}. As per our readers, the highest difference in the quality of segmentation between the Freesurfer segmentation and Finetuned model segmentation is seen for those regions which showed major boundary errors in the Freesurfer segmentation as demonstrated in figure \ref{fig:faulty_seg_insula}. Those regions are Insula and Pallidum. 

In the Freesurfer segmentation, we see boundary errors, inclusion of discontiguous areas, as well as stair-step artifacts along the boundary that render a noisy and non-natural-appearing result. One potential benefit of a deep-learning-based brain segmentation tool over the traditional MRF-based tool is that by training over multiple examples, the model learns that jagged or stair-step boundaries are not consistent, and can not be explained by naturally visible MRI boundaries. The model therefore simply fails to learn the arbitrary jaggedness. For additional examples of Freesurfer vs DenseUNet putamen and pallidum segmentations please refer to figure \ref{fig:faulty_seg_putamen} and figure \ref{fig:faulty_seg_pallidum} in the appendix. 


\subsection{Limited high quality manual data for quantitative evaluation and finetuning}
As we see in section \ref{sec:Freesurfer errors} and from the results in table \ref{table:reader_study_results}, the Freesurfer generated segmentation labels have low quality and hence, the model trained only on such labels would also be susceptible to generate low quality segmentation. A manually annotated high quality segmentation, therefore, becomes an important resource for training/finetuning the model and quantatively evaluating its performance.

\begin{figure}[ht]
\begin{subfigure}{.5\textwidth}
  \centering
  \includegraphics[width=.8\linewidth]{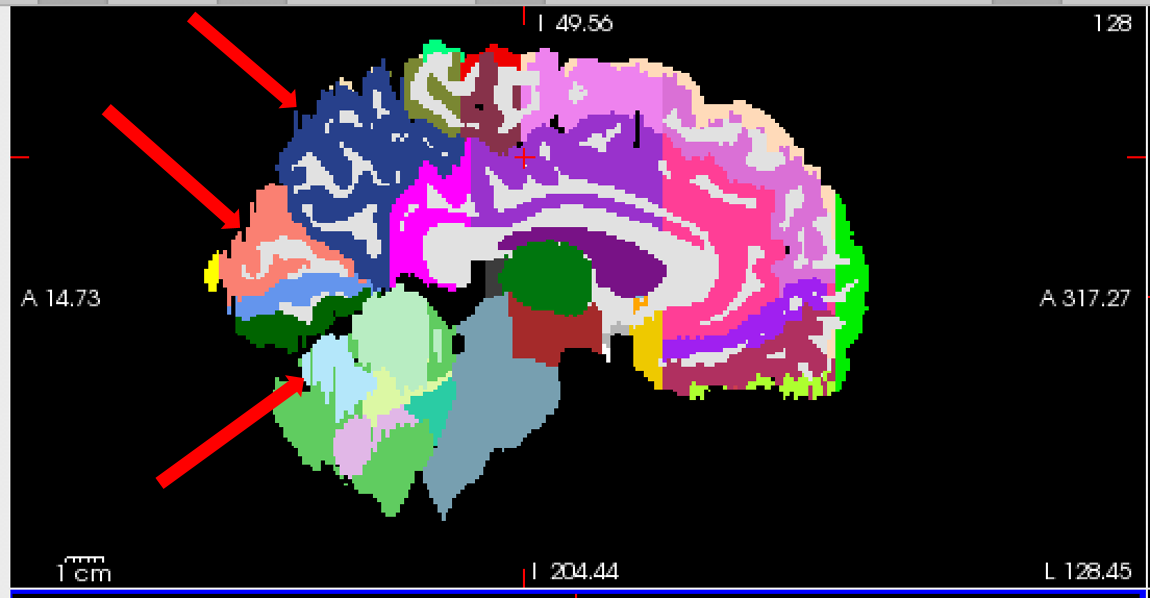}
  \caption{Sagittal View}
  \label{fig:trace_back_artifacts_1}
\end{subfigure}%
\begin{subfigure}{.5\textwidth}
  \centering
  \includegraphics[width=.8\linewidth]{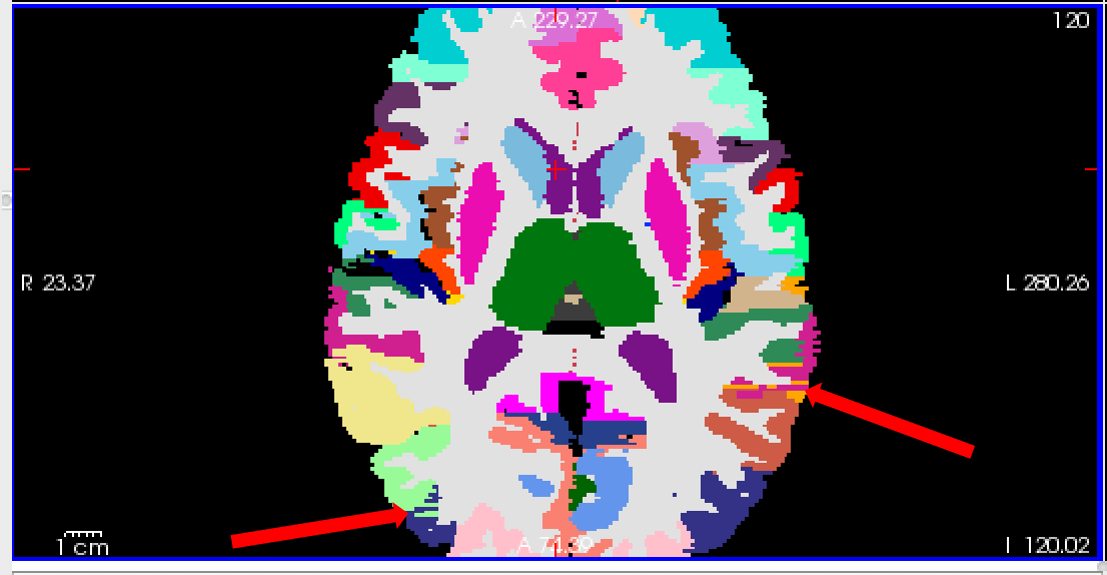}
  \caption{Axial View}
  \label{fig:trace_back_artifacts_2}
\end{subfigure}
\caption{Image showcasing artifacts induced by performing manual correction in one view (here coronal) and extending to other views (here sagittal and axial)}
\label{fig:trace_back_artifacts}
\end{figure}

There are two major open source datasets which contain manually annotated complete brain segmentations. Most prior work in the area of developing deep-learning-based brain segmentation models use the 2012 MICCAI Multi-Atlas labelling challenge dataset \cite{2012miccai}. The ground truth labels of this dataset when visualized, show corruption in sagittal and axial views. 
This is due to the manual segmentation being performed only in coronal view without correction in other planes. These artifacts, visualized in the figure \ref{fig:trace_back_artifacts}, introduce potential unwanted biases in models that train on these data. 

Because of these problems with the manual segmentation ground truth observed in the 2012 MICCAI Multi-Atlas labelling challenge dataset, we use the Mindboggle 101 dataset which contains 102, manually segmented regions in each brain MRI. Mindboggle 101 has the benefit of containing image data from a diverse set of scanners and sites (discussed in the dataset section) which gives our model the opportunity to learn the invariances across the image protocols and scanners, making a more robust and generalizable model.

\subsection{Limitations of this study} \label{sec:limitations}
Limitations of this study include:
\begin{enumerate}
    \item At the current time, the model offers segmentation of gray matter sub-structures (aparc) and does not include white matter sub-structure segmentation (wmparc), though this is an ongoing part of our future work.
    \item Some regions such as white matter hyperintensities, which are not commonly represented in the HCP dataset, were excluded. 
    \item In the neuroimaging community, there are different approaches to segmentation. This work adheres closely to the DKT protocol followed by Mindboggle and Freesurfer \cite{DKT}; other segmentation protocols would require re-training of the model. 
    \item More extensive reader study involving the full 102 regions of interest was not present in this work and is deferred to future studies. 
\end{enumerate}


\section{Future Work and Conclusion}
We have a number of active directions currently under investigation: Firstly,as discussed in section \ref{sec:limitations}, we have not validated white matter regional segmentation, which is underway. Secondly, lately, there is much interest to learn from noisy labels. Since the first step of this model involves learning from auxiliary and somewhat noisy Freesurfer labels, employing additional methods to more efficiently learn from noisy labels could be explored. Lastly, we are exploring ways of selectively fine tuning the model only for those regions with high quality manual segmentation available.

In conclusion, we present a deep-learning based model that performs an extensive, anatomical brain segmentation yielding 102 brain regions that match a commonly used tool, Freesurfer, in one minute (for a single GPU machine) for a single brain volume. This fast and accurate open-source segmentation tool can finally make on-demand clinical utilization of brain segmentation feasible, enabling translation of wealth of neurological research into clinic. Our proposed model does not need a complete 3D Brain MR for performing the segmentation. Even a single slice of Brain T1 MRI could be segmented using our proposed model, and the model operates without any additional pre- or post-processing steps. We demonstrate successful generalization of our model to a variety of scanner devices and resolutions. Finally, we also performed first reader study to evaluate the segmentation quality of Freesurfer and proposed model and show that the proposed model's segmentations are of superior quality compared to Freesurfer.

\section{Open Source Tool}
Training and inference code, a jupyter notebook for full instructions, and pre-trained models including Coronal, Axial and Sagittal finetuned and non-finetuned models can be accessed from \url{https://github.com/NYUMedML/DARTS}.  

\bibliographystyle{plain}
\bibliography{template}

\section*{Appendix}
\subsection*{Schematic diagram for U-Net model}
\begin{figure}[ht]
    \centering
    \includegraphics[width=\linewidth]{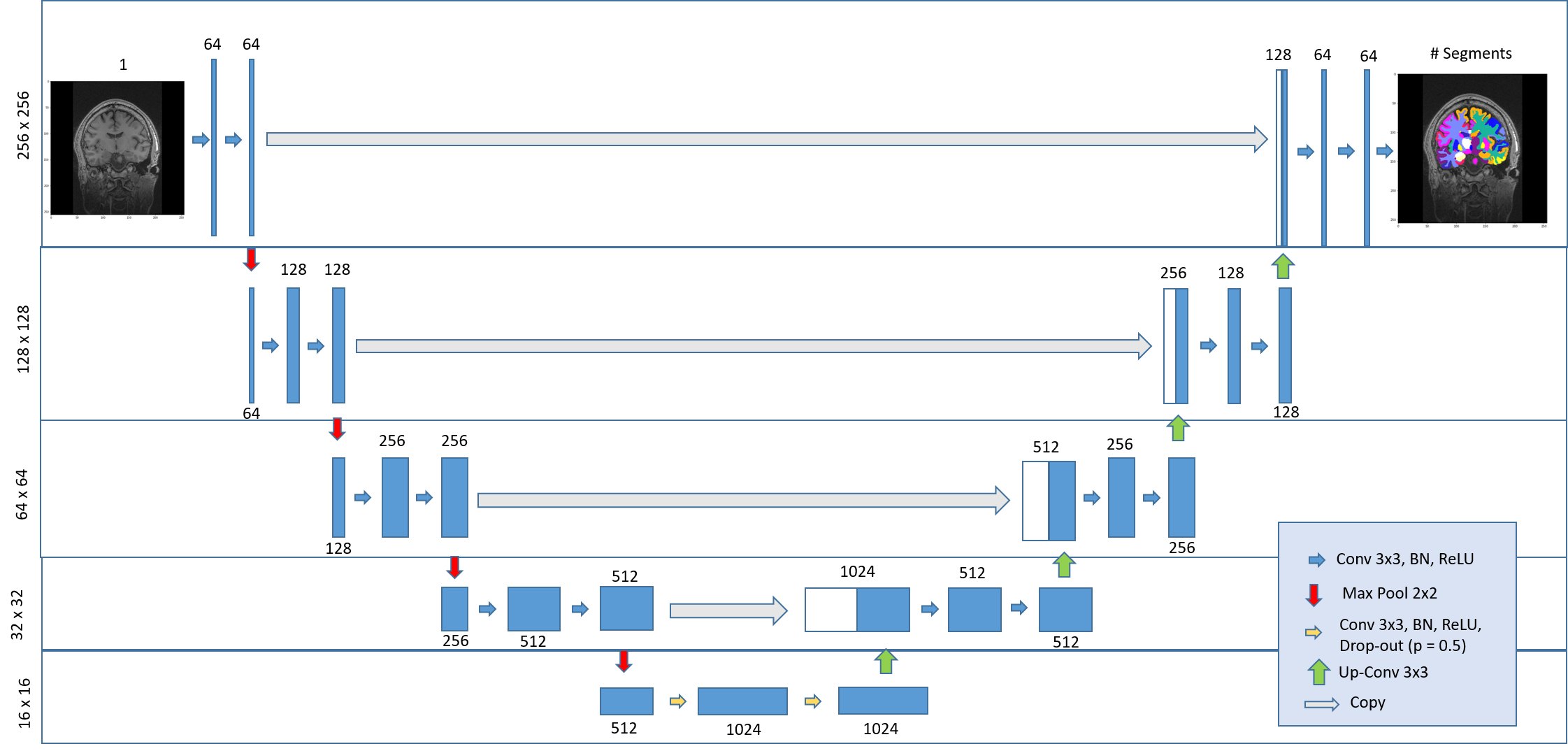}
    \caption{Schematic Diagram of Vanilla U-Net}
    \label{fig:unet}
\end{figure}

Figure \ref{fig:unet} shows the schematic diagram for the vanilla U-Net model.
\subsection*{Plots for Voxel Count}
\begin{figure}[t]
    \centering
    \includegraphics[width=\linewidth]{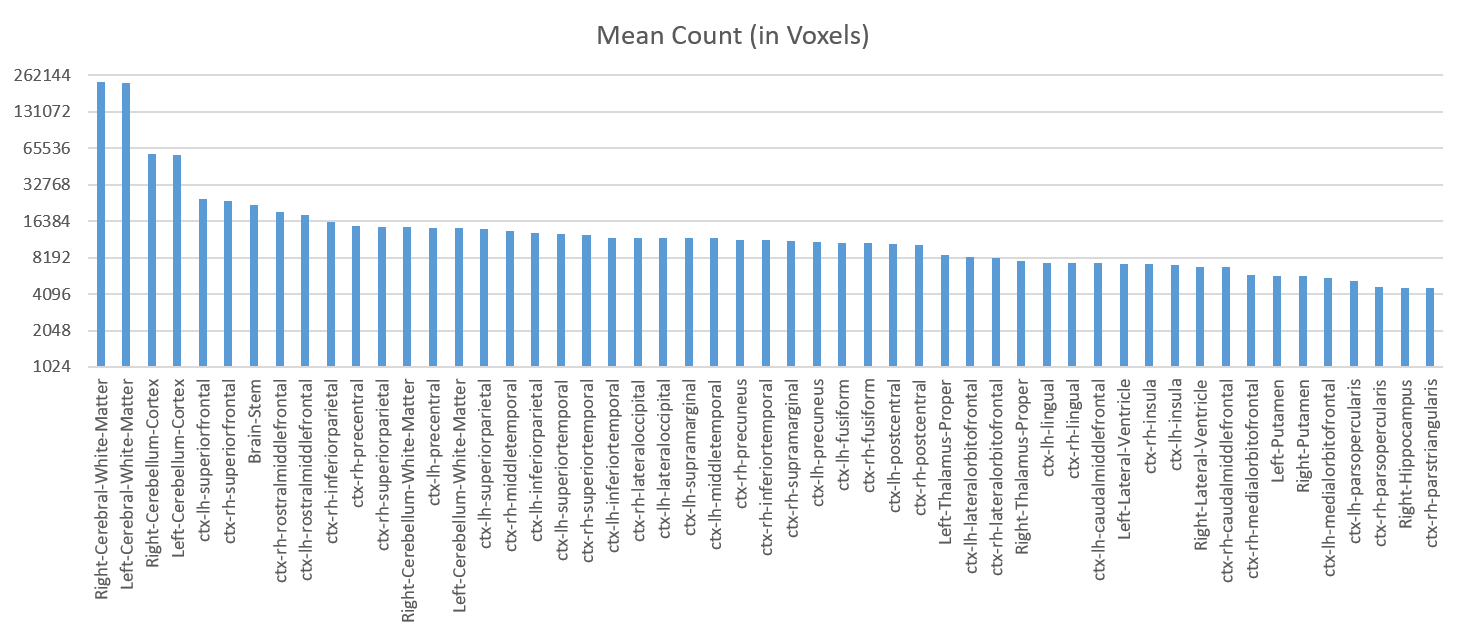}
    \caption{Plot of mean voxel count for top 53 Region of Interests (ROIs): The count is taken over 20 datasets and the average is reported. Average count stands for the average number of voxels for a particular label in the Freesurfer output}
    \label{fig:voxel_count_top}
\end{figure}

\begin{figure}[t]
    \centering
    \includegraphics[width=\linewidth]{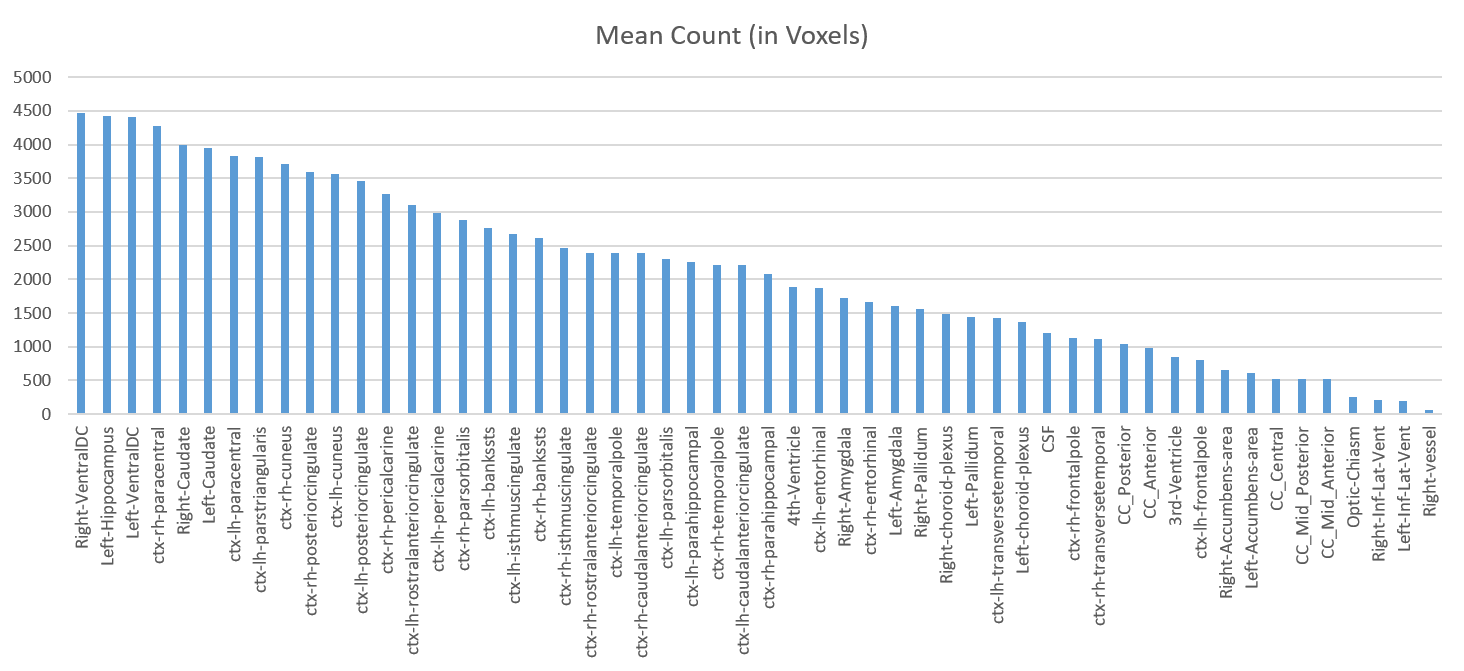}
    \caption{Plot of mean voxel count for bottom 54 Region of Interests (ROIs): The count is taken over 20 datasets and the average is reported. Average count stands for the average number of voxels for a particular label in the Freesurfer output}
    \label{fig:voxel_count_bottom}
\end{figure}

Figure \ref{fig:voxel_count_top} and figure \ref{fig:voxel_count_bottom} shows the mean voxel count for each class in an MRI.

\subsection*{Detailed Dice scores for all the ROIs for DenseUNet model}
\subsubsection*{NYU Data}
\begin{figure}[ht]
    \centering
    \includegraphics[width=0.9\linewidth]{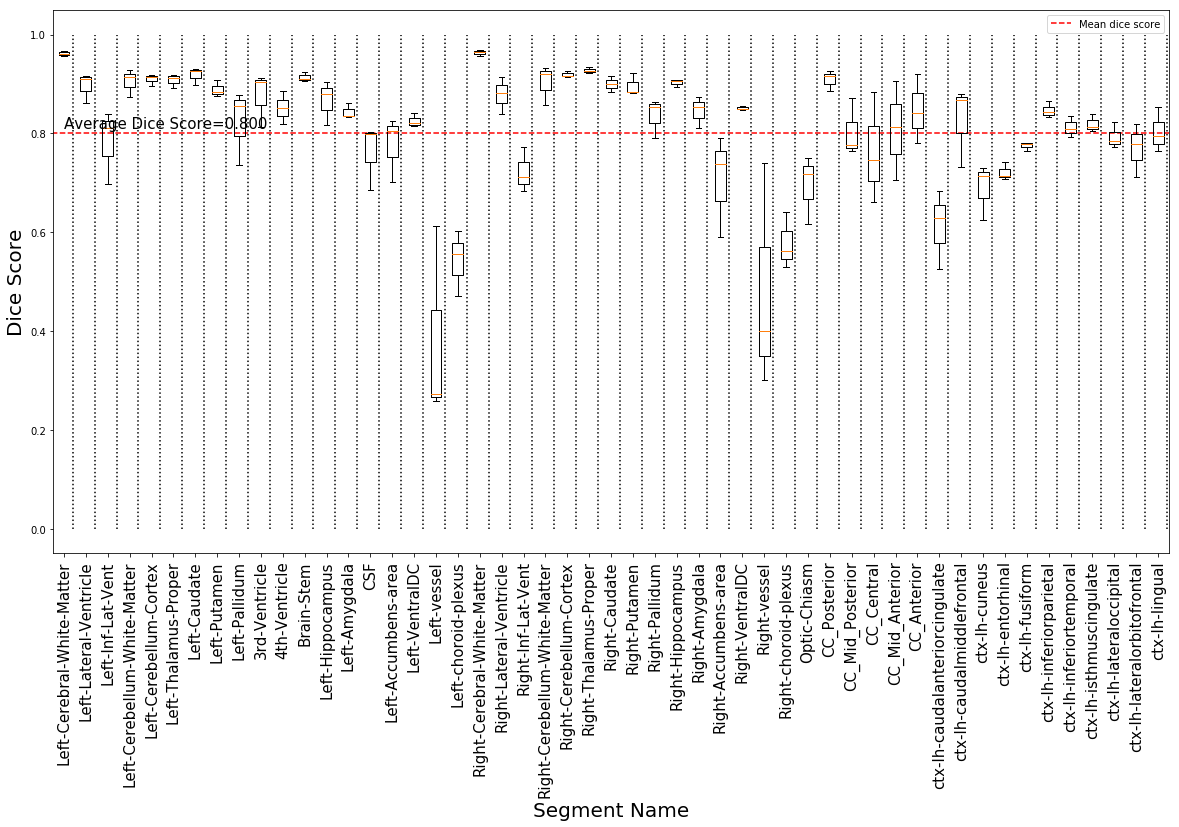}
    \caption{Box plot showing the detailed dice scores for the First 51 ROIs of NYU Dataset }
    \label{fig:dice_plot_nyu_part_1}
\end{figure}

\begin{figure}[ht]
    \centering
    \includegraphics[width=0.9\linewidth]{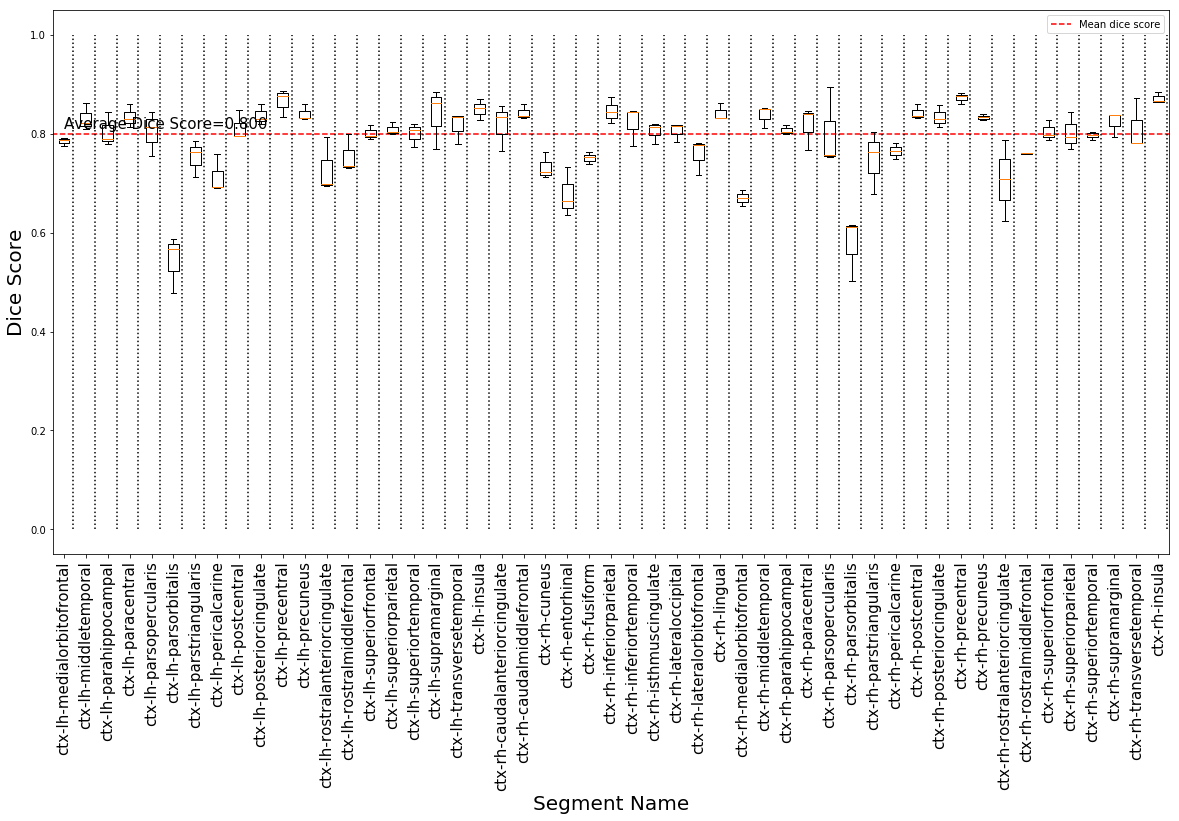}
    \caption{Box plot showing the detailed dice scores for the last 51 ROIs of NYU Dataset }
    \label{fig:dice_plot_nyu_part_2}
\end{figure}

The plots \ref{fig:dice_plot_nyu_part_1} and \ref{fig:dice_plot_nyu_part_2} show the box-plot of dice scores for all the ROIs for NYU dataset.

\subsection*{Comparison of Finetuned v/s non-Finetuned model}
\subsubsection*{DenseUNet}
\begin{figure}[ht]
    \centering
    \includegraphics[width=0.9\linewidth]{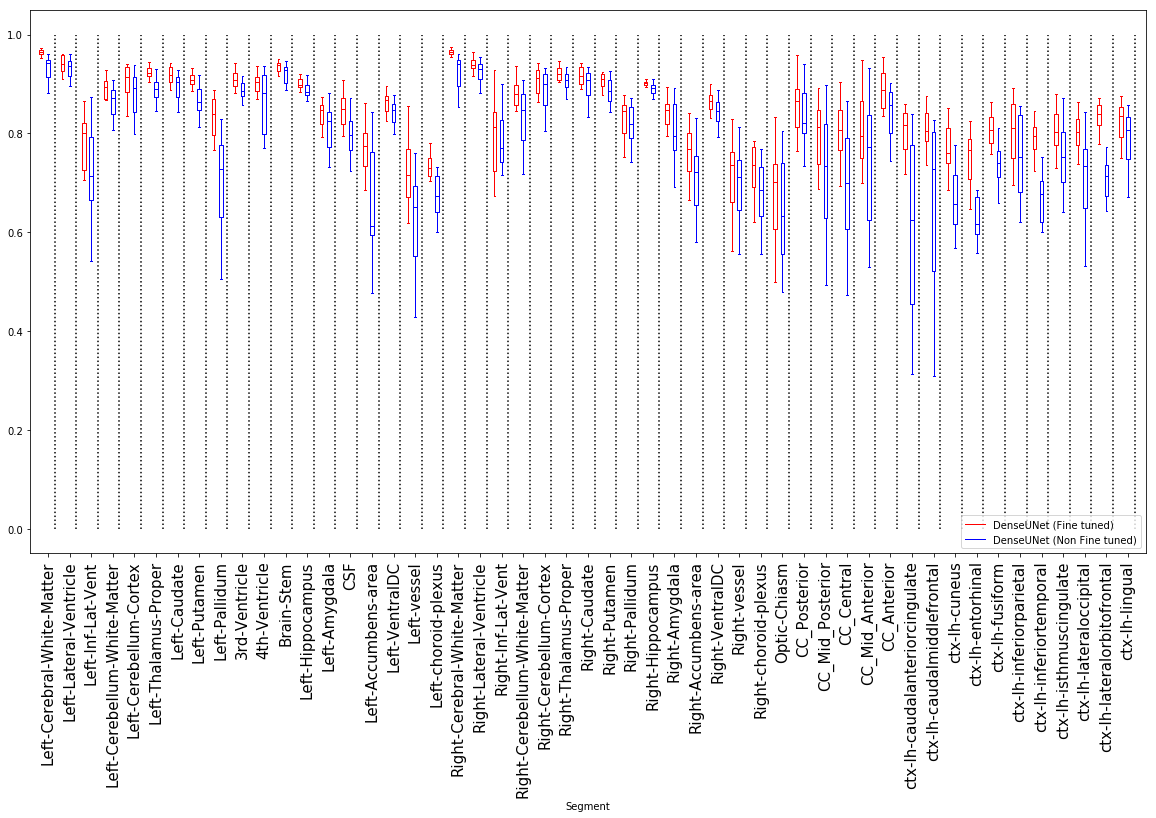}
    \caption{Box plot showing the comparison of dice scores for the Finetuned and non-Finetuned DenseUNet (First 51 ROIs of Mindboggle dataset) }
    \label{fig:compare_dice_plot_man_part_1}
\end{figure}

\begin{figure}[ht]
    \centering
    \includegraphics[width=0.9\linewidth]{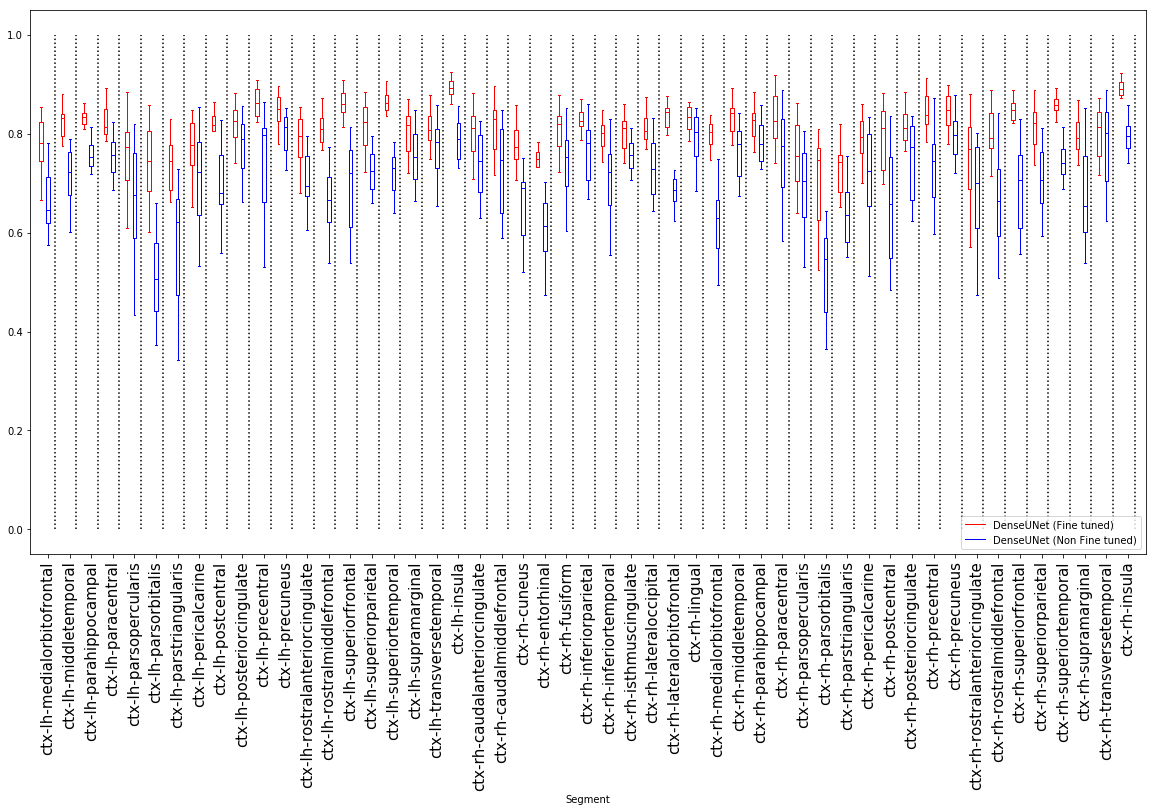}
    \caption{Box plot showing the comparison of dice scores for the Finetuned and non-Finetuned DenseUNet (Last 51 ROIs of Mindboggle dataset) }
    \label{fig:compare_dice_plot_man_part_2}
\end{figure}

The plots \ref{fig:compare_dice_plot_man_part_1} and \ref{fig:compare_dice_plot_man_part_2} show the difference in the model's performance on the Mindboggle Data with and without finetuning.

\subsection*{Comparison of DenseUNet v/s U-Net model}
\begin{figure}[ht]
    \centering
    \includegraphics[width=0.9\linewidth]{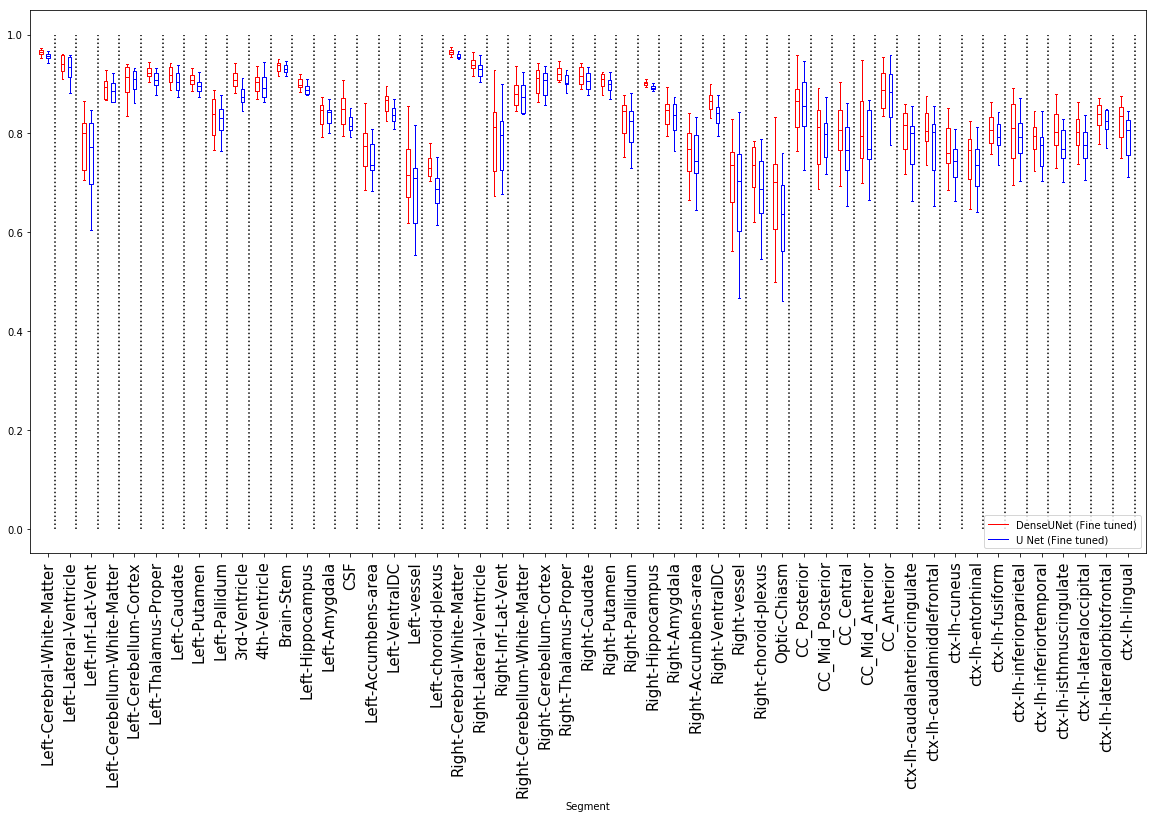}
    \caption{Box plot showing the comparison of dice scores for DenseUNet and U-Net model(First 51 ROIs of Mindboggle dataset) }
    \label{fig:compare_dice_plot_man_dnu_part_1}
\end{figure}

\begin{figure}[ht]
    \centering
    \includegraphics[width=0.9\linewidth]{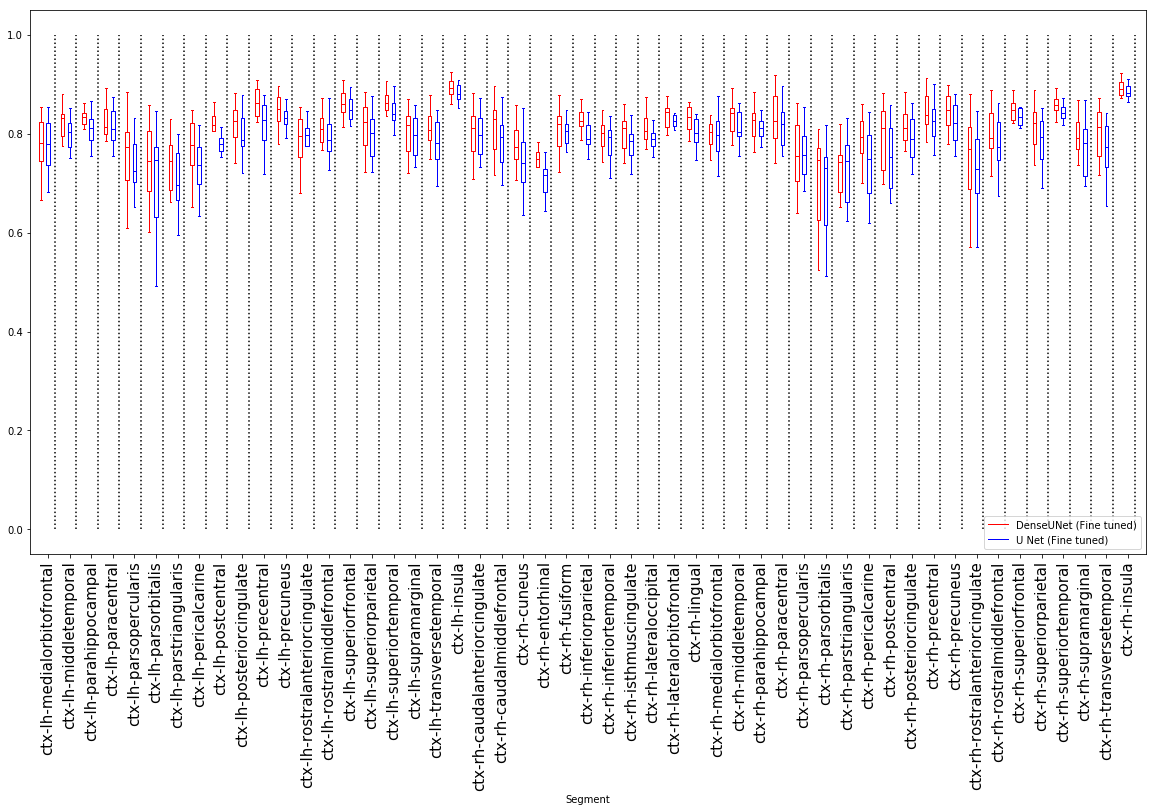}
    \caption{Box plot showing the comparison of dice scores for DenseUNet and U-Net model(Last 51 ROIs of Mindboggle dataset) }
    \label{fig:compare_dice_plot_man_dnu_part_2}
\end{figure}

The plots \ref{fig:compare_dice_plot_man_dnu_part_1} and \ref{fig:compare_dice_plot_man_dnu_part_2} show the difference in the model's performance on the Mindboggle Data for DenseUNet and U-Net.

\subsection*{Faulty Freesurfer Segmentation}
\subsubsection*{Putamen}
\begin{figure}[ht]
    \centering
    \includegraphics[width=0.6\linewidth]{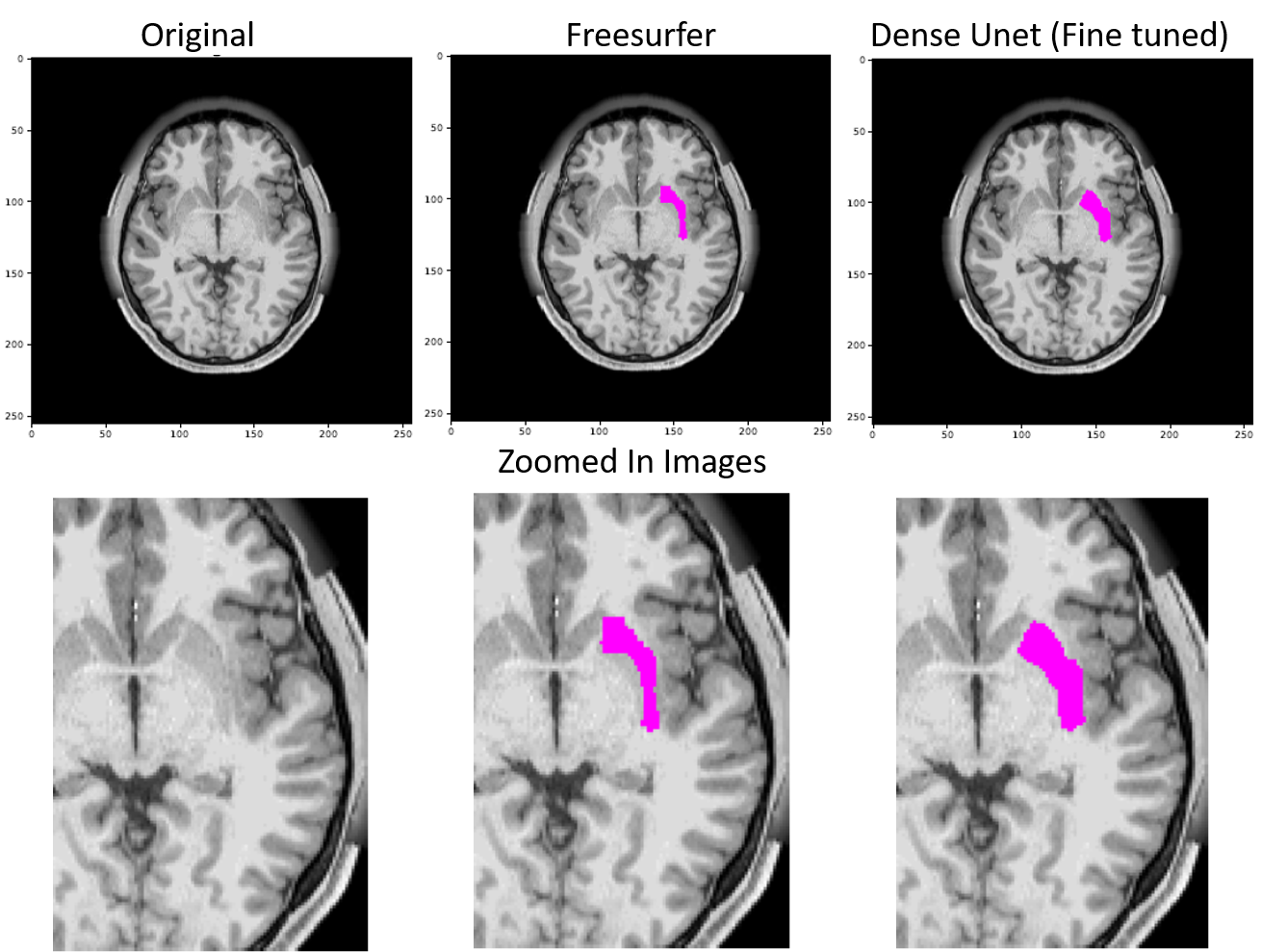}
    \caption{The image shows the original slice of an MRI, Freesurfer's (FS's) prediction for Putamen and the proposed model's prediction of the same. It is evident that FS's prediction don't obey the boundaries and are non-natural looking grainy segmentation whereas the proposed model's prediction obey the segment boundaries and are much more natural looking }
    \label{fig:faulty_seg_putamen}
\end{figure}

Figure \ref{fig:faulty_seg_putamen} shows the difference in the segmentation outputs of Freesurfer and the proposed model.

\subsubsection*{Pallidum}
\begin{figure}[ht]
    \centering
    \includegraphics[width=0.6\linewidth]{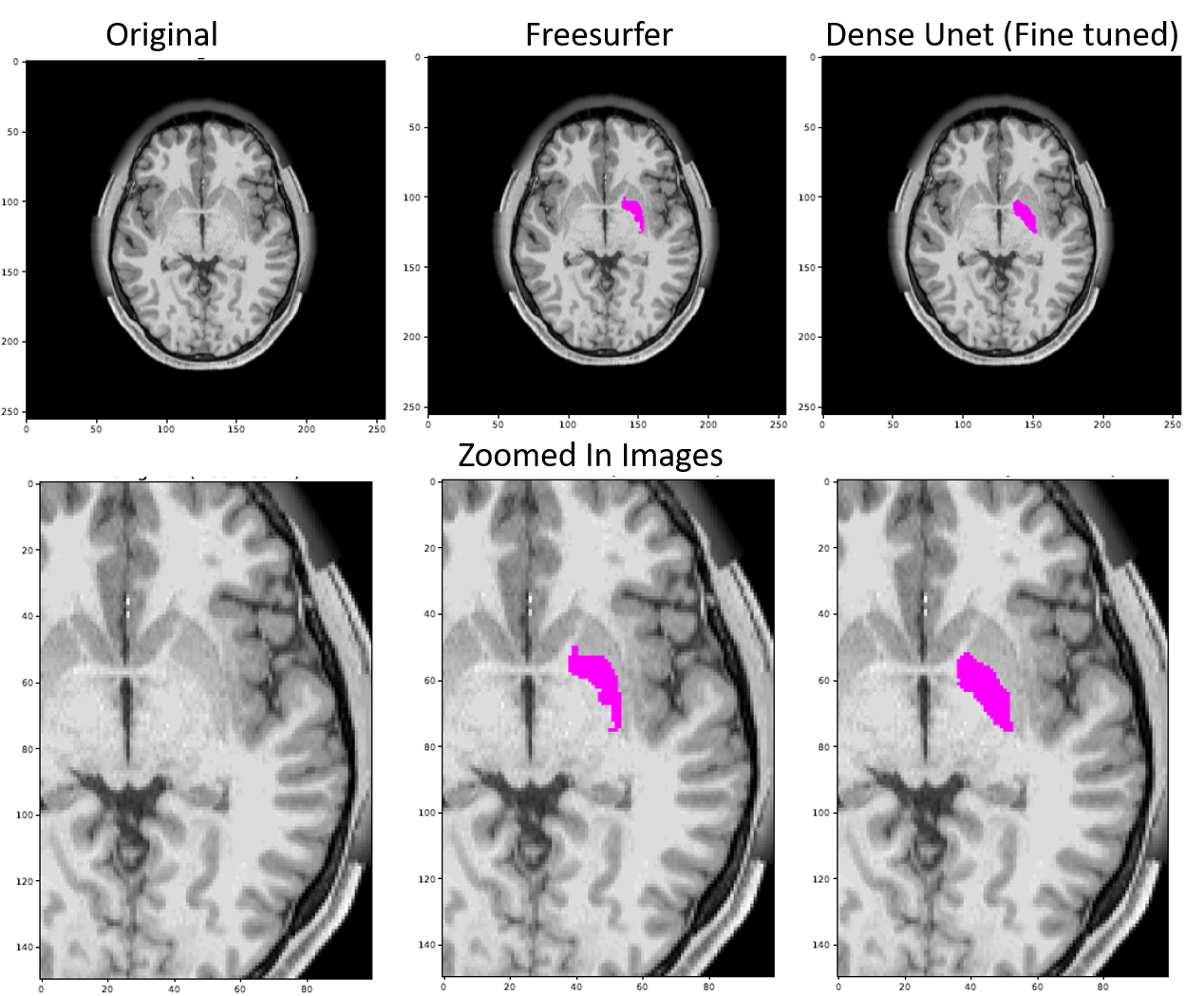}
    \caption{The image shows the original slice of an MRI, Freesurfer's (FS's) prediction for Pallidum and the proposed model's prediction of the same. It is evident that FS's prediction don't obey the boundaries and are non-natural looking grainy segmentation whereas the proposed model's prediction obey the segment boundaries and are much more natural looking }
    \label{fig:faulty_seg_pallidum}
\end{figure}

Figure \ref{fig:faulty_seg_pallidum} shows the difference in the segmentation outputs of Freesurfer and the proposed model.

\subsection*{Comparison of models trained from different views}
\begin{figure}[ht]
    \centering
    \includegraphics[width=0.9\linewidth]{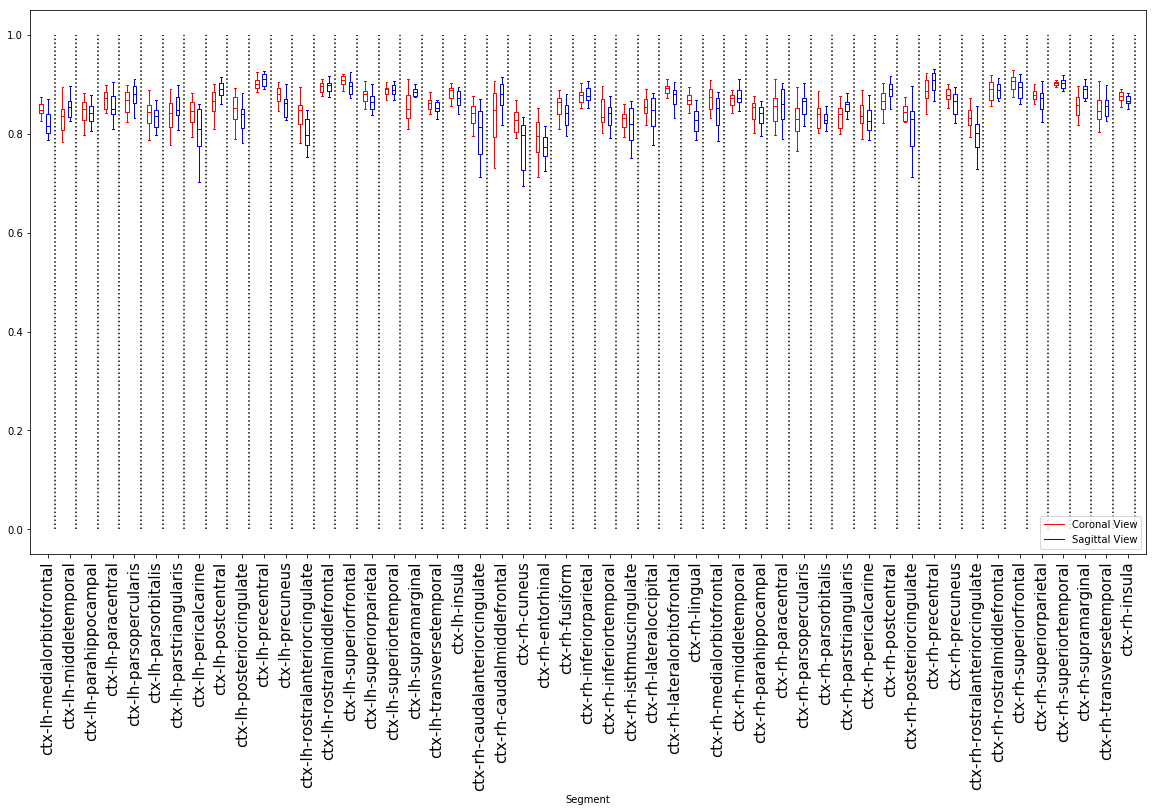}
    \caption{Box plot showing the comparison of dice scores for the coronally and sagittally trained model (First 51 ROIs of HCP data) }
    \label{fig:compare_dice_plot_aparc_hcp_cvs_part_1}
\end{figure}

\begin{figure}[ht]
    \centering
    \includegraphics[width=0.9\linewidth]{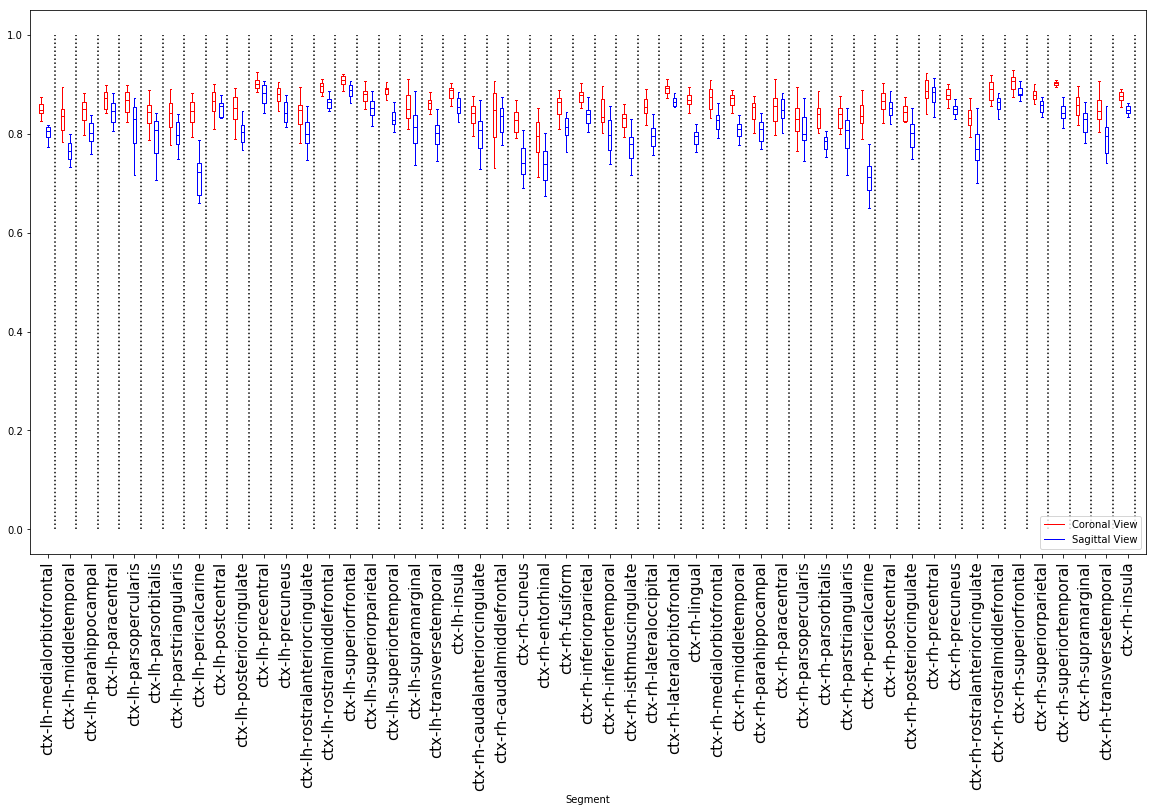}
    \caption{Box plot showing the comparison of dice scores for the coronally and sagittally trained model (Last 51 ROIs of HCP data) }
    \label{fig:compare_dice_plot_aparc_hcp_cvs_part_2}
\end{figure}

\begin{figure}[ht]
    \centering
    \includegraphics[width=0.9\linewidth]{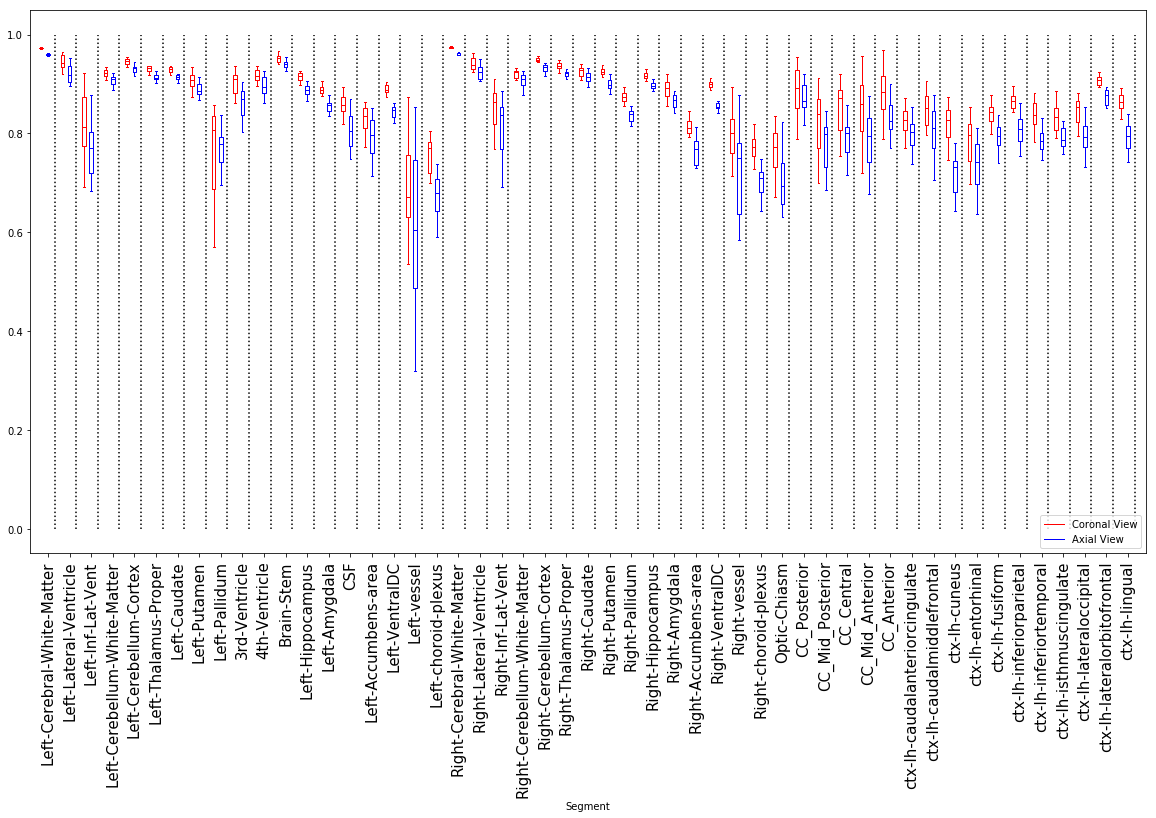}
    \caption{Box plot showing the comparison of dice scores for the coronally and axially trained model (First 51 ROIs of HCP data) }
    \label{fig:compare_dice_plot_aparc_hcp_cva_part_1}
\end{figure}

\begin{figure}[ht]
    \centering
    \includegraphics[width=0.9\linewidth]{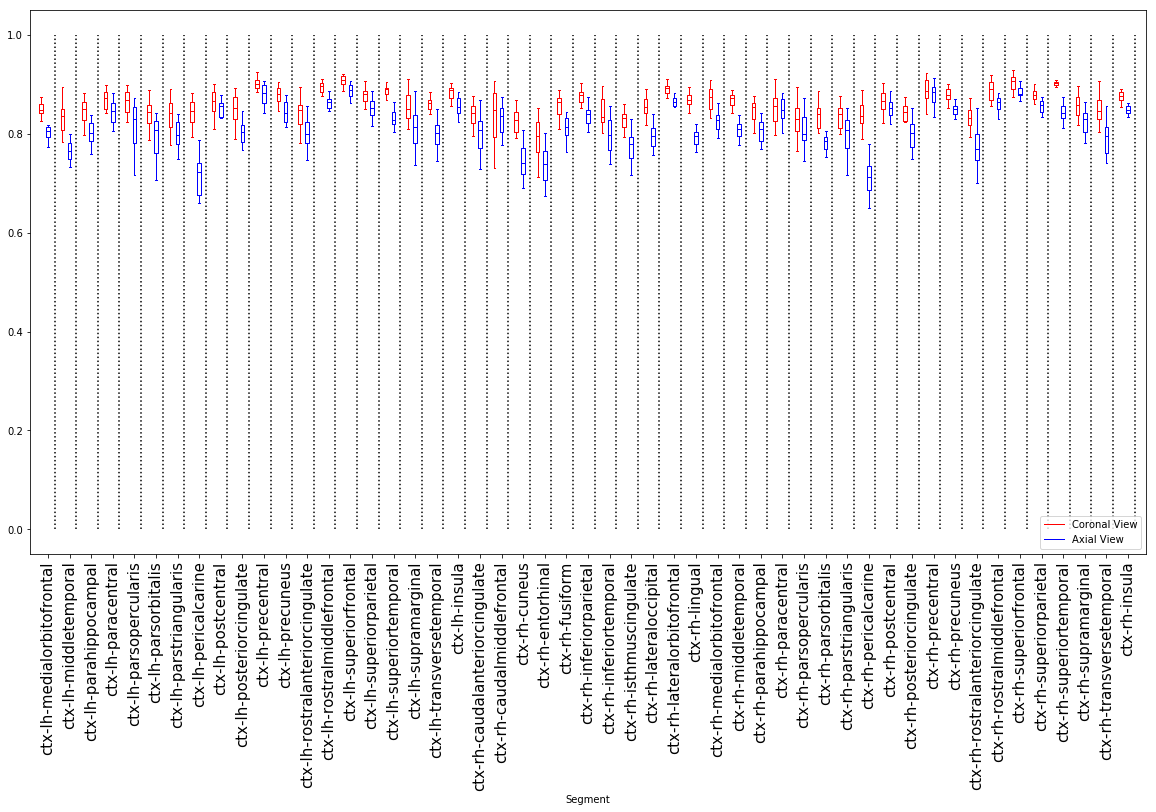}
    \caption{Box plot showing the comparison of dice scores for the coronally and axially trained model (Last 51 ROIs of HCP data) }
    \label{fig:compare_dice_plot_aparc_hcp_cva_part_2}
\end{figure}

Plots \ref{fig:compare_dice_plot_aparc_hcp_cva_part_1}, \ref{fig:compare_dice_plot_aparc_hcp_cva_part_2}, \ref{fig:compare_dice_plot_aparc_hcp_cvs_part_1}, and \ref{fig:compare_dice_plot_aparc_hcp_cvs_part_2} show the difference in the performance of the model trained using 2D slices from different views (coronal, sagittal and axial). For performing the initial view selection, we train the models only using 30,000 2D slices of HCP data and validated using 5,000 2D slices. The validation set is also from the HCP dataset. Based on the preliminary experiment, it was seen that model trained using the 2D coronal slices performed the best for all the ROIs.

\subsection*{Correlation of dice score with the size of ROI}
\begin{figure}[ht]
    \centering
    \includegraphics[width=0.6\linewidth]{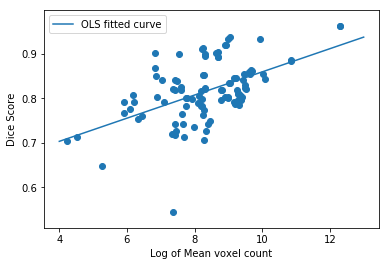}
    \caption{Plot shows the dice score vs log(mean voxel count) for all 102 ROIs (each ROI is a represented as a point on the plot). Pearson's correlation coefficient = 0.521 and is statistically significant with 99.9\% significance level.}
    \label{fig:dice_count_reg_plot}
\end{figure}

The plot \ref{fig:dice_count_reg_plot} shows the variation of dice score with the size of the ROI.

\end{document}